\documentclass{article}
\IfFileExists{arxiv.sty}{\usepackage{arxiv}}{%
  \usepackage[margin=1in]{geometry}
  \usepackage{fancyhdr}
  \pagestyle{fancy}\fancyhf{}\fancyfoot[C]{\thepage}
  
}
\usepackage[utf8]{inputenc} 
\usepackage[T1]{fontenc}    
\usepackage{hyperref}       
\usepackage{url}            
\usepackage{booktabs}       
\usepackage{amsfonts}       
\usepackage{nicefrac}       
\usepackage{microtype}      
\usepackage{graphicx}
\usepackage[authoryear,round]{natbib}
\usepackage{bm}             
\usepackage{dcolumn}        
\usepackage{amsmath,amssymb}
\usepackage{adjustbox}
\usepackage{algorithm,algpseudocode}
\usepackage{placeins}
\hypersetup{
    colorlinks = true,
    urlcolor   = blue,
    citecolor  = black,
    linkcolor  = black,
}
\title{BridgeSRDA: Efficient Probabilistic Super-Resolution Data Assimilation with Forecast-Informed Stochastic Interpolants}

\author{
A. S. P. Ayapilla$^{1,*}$, 
Y. Yasuda$^{2}$, 
R. Onishi$^{1}$\\
\small
$^{1}$Supercomputing Research Center, Institute of Integrated Research,\\
Institute of Science Tokyo, Tokyo 152-8550, Japan\\
$^{2}$Research Institute for Value-Added-Information Generation, JAMSTEC,\\
Yokohama, Japan\\
$^{*}$Corresponding author: ayapilla.asp@scrc.iir.isct.ac.jp
}
\date{Working draft: October 13, 2026}
\begin{document}\maketitle
\begin{abstract}
High-resolution ensemble data assimilation is computationally demanding because it requires repeated forecasts and updates of many high-dimensional flow states. We introduce BridgeSRDA, a probabilistic super-resolution data assimilation (SRDA) framework that combines low-resolution numerical forecasts with sparse, noisy observations to generate high-resolution analysis windows. A neural network learns a forecast-informed stochastic transport, which is sampled to produce an ensemble of reconstructions. Each current analysis initializes the next low-resolution forecast, while the same window provides retrospective reconstructions and short-range nowcasts. We evaluate BridgeSRDA over ten independently initialized trajectories of a three-dimensional quasi-geostrophic flow in a narrow-jet regime different from the training configuration. At the observation spacing used during training, the dense-observation variant approaches the mean and structural reconstruction accuracy of a high-resolution local ensemble transform Kalman filter and reduces the mean absolute error ratio by approximately 32.5\% relative to a reference deterministic four-dimensional SRDA model. With 30 ensemble members, BridgeSRDA requires approximately 17 s per analysis, compared with 31 s for the filter under the recorded execution conditions. Reducing the ensemble to ten members lowers the cost to approximately 6.6 s with little change in mean reconstruction accuracy, although fine-scale accuracy and interval coverage are affected. Additional experiments show that execution improvements and less frequent analysis updates can reduce cycling costs further. BridgeSRDA achieves a competitive continuous ranked probability score, but its ensembles are underdispersed: nominal 90\% intervals contain approximately 62\% of reference values. Boundary spread artifacts and reduced skill under sparser observations also warrant further attention. These results demonstrate the potential of forecast-informed stochastic transport for accurate and computationally flexible probabilistic SRDA, while identifying uncertainty calibration and adaptation to changing observation networks as priorities for future development.
Source code for this work is available at \url{https://github.com/Pranithhazard/BridgeSRDA}.
\end{abstract}

\FloatBarrier
\section{Introduction}\label{sec:introduction}

Accurate weather forecasts depend on a reliable estimate of the current state of the atmosphere. Observations from satellites, surface stations and other instruments provide valuable information about that state, but they are incomplete and affected by measurement errors. Data assimilation (DA) combines these observations with a short-range forecast to produce an updated estimate, called the analysis, which initializes the next forecast. Ensemble methods represent uncertainty through a collection of possible states and use their differences to estimate how forecast errors vary across the system. Variational methods instead seek an analysis that balances departures from the forecast and observations according to their error statistics. These approaches form the basis of operational forecasting systems, often in combination. For example, the Japan Meteorological Agency (JMA) uses ensemble and variational methods in its global and mesoscale systems, while the European Centre for Medium-Range Weather Forecasts (ECMWF) uses an ensemble of variational assimilation systems to represent analysis uncertainty~\citep{Carrassi2018Overview,JMA2026NWP,ECMWF2026DA}. As the resolution and complexity of a forecast system increase, repeatedly producing analyses and advancing ensembles of forecasts can become computationally expensive.

Machine learning (ML) offers several ways to reduce the cost of forecasting and state estimation. Learned forecast models such as GraphCast and GenCast have demonstrated skillful global medium-range prediction, while Aardvark Weather extends learning to the observation-to-forecast pipeline~\citep{Lam2023GraphCast,Price2025GenCast,Allen2025Aardvark}. Applying learned forecast models to other settings requires careful assessment of how errors develop over successive predictions and how well the learned dynamics generalize to different scales and flow conditions~\citep{Bonavita2024Limitations,Duraisamy2026Surrogates}. An alternative is to retain a numerical forecast model and use ML to recover fine-scale information from a less expensive coarse simulation. This approach can build on existing solvers and their representation of the resolved dynamics.

Super-resolution (SR) methods provide a means of reconstructing high-resolution (HR) fields from low-resolution (LR) information. In urban micrometeorology, \citet{Onishi2019UrbanSR} combined numerical weather simulations with learned SR to make fine-scale prediction more computationally accessible. Subsequent work extended this approach to three-dimensional temperature and velocity fields around buildings and to two-stage reconstruction of street-scale atmospheric flows~\citep{Yasuda2023Urban,Yasuda2025TwoStage}. SR has also been applied to turbulent flows, including the reconstruction of flow evolution in both space and time~\citep{Fukami2021Spatiotemporal}; a broader survey of fluid-flow applications is given by \citet{Fukami2023SRSurvey}. In weather forecasting, learned downscaling has been used to generate high-resolution probabilistic precipitation forecasts~\citep[and references therein]{Harris2022Precipitation}. Super-resolution data assimilation (SRDA) brings this capability together with observations to produce HR analyses while retaining an LR numerical forecast. Several formulations have been explored. \citet{Barthelemy2022} combined learned SR with an ensemble Kalman filter (EnKF) update and returned the analysis to the coarse grid for the next forecast. \citet{Yasuda2023SRDA} developed a spatiotemporal SRDA method that directly combines LR forecasts and observations to reconstruct a window of HR states, and \citet{Notsu2025} extended this approach to three-dimensional quasi-geostrophic (QG) flow. Probabilistic formulations have also been investigated. In particular, \citet{Yasuda2025Unsupervised} developed an unsupervised SRDA method based on a conditional variational autoencoder and showed how the SR operator can contribute to the representation of background-error covariances.

Generative models provide another way to produce multiple HR states consistent with available forecast and observation information. Diffusion models have been used for weather-scale DA and for reconstructing flow fields from incomplete information~\citep{Rozet2023SDA,Huang2024DiffDA,Li2024SparseDynamics,Du2024CoNFiLD}. In our previous study, DiffSRDA, we used a conditional denoising diffusion probabilistic model (DDPM) to generate HR analysis windows from LR forecasts and sparse observations for a two-dimensional barotropic jet~\citep{Ayapilla2026DiffSRDA}. DiffSRDA approached the reconstruction accuracy of an HR-driven EnKF, improved on deterministic neural SRDA and produced ensemble spread concentrated around dynamically active flow structures. We also investigated how observations could guide the model when sensor layouts differed from those used during training. These results motivate the use of generative models for probabilistic SRDA and raise the question of how well such methods perform in higher-dimensional systems.

Here we investigate that question using three-dimensional QG jet turbulence. The system remains a reduced model of balanced flow, but its vertical coupling and evolving small-scale structures make it a more demanding reconstruction problem than the two-dimensional barotropic jet. Each nine-frame HR window in the present study contains 1,179,648 potential-vorticity (PV) values, compared with 40,960 values in the five-frame HR windows used in DiffSRDA. The larger state and temporal window increase the burden of both training and repeated ensemble generation. A DDPM-style conditional diffusion model adapted from our previous work did not produce useful reconstructions under the configurations examined here. This experience led us to reconsider how the generative model could make use of the information already present in the LR forecast.

In the conditional DDPM used in DiffSRDA, the LR forecast and observations guide the generation, but sampling begins from Gaussian noise. The LR forecast itself already provides an organized, dynamically evolved approximation to the flow. Starting from this forecast could allow the generative model to focus its transport on the differences between the coarse forecast and the desired HR analysis. Stochastic interpolants provide a flexible framework for constructing such transformations between prescribed source and target distributions~\citep{Albergo2023Interpolants,Lipman2023FlowMatching}. Related work supports exploring informative sources for flow reconstruction. \citet{Yasuda2026Bridge} used a Schr\"odinger-bridge-based model to reconstruct urban near-surface temperature from coarse numerical fields and obtained accuracy comparable to a diffusion model with fewer neural evaluations. \citet{Schiodt2026TurbulenceSI} likewise used stochastic interpolants to generate super-resolved turbulent fields from coarse flow information. These studies motivate the use of forecast-informed transport for probabilistic reconstruction. Sequential DA adds the further requirements that generated states incorporate observations, provide useful estimates of uncertainty and initialize subsequent forecasts.

Motivated by these considerations, we develop BridgeSRDA, a probabilistic spatiotemporal SRDA framework that combines LR numerical forecasting with observation-conditioned stochastic transport. Here, ``bridge'' refers to the stochastic transport connecting the forecast-informed source distribution to the high-resolution target distribution. At each cycle, a neural model transforms a lifted LR forecast into an ensemble of HR windows using the available observations. Each window contains reconstructions of earlier states, an analysis at the current time and short-range nowcasts. The current analysis of each ensemble member initializes its next LR forecast, allowing us to examine how reconstruction errors and uncertainty evolve over repeated cycles. The retrospective reconstructions use observations available later than the states being estimated and therefore have a smoothing role, while the current analysis has a filtering role. Related generative approaches have also explored the reconstruction of individual states and trajectories from incomplete observations, including score-based DA and flow-based atmospheric assimilation~\citep{Rozet2023SDA,Chakraborty2026FlowDA}. A broader perspective on the connections between ML and DA is provided by \citet{Arcucci2026Convergence}.

Our objective is to assess whether forecast-informed stochastic transport can support accurate, uncertainty-aware and computationally efficient sequential DA when combined with numerical forecasting. We develop BridgeSRDA as an initial framework for investigating this question and compare it with deterministic SRDA and ensemble-based assimilation methods over repeated cycles. Our evaluation considers reconstruction accuracy, fine-scale structure, ensemble spread and calibration, together with the effects of observation representation, observation sparsity and sampling budget. These experiments establish the capabilities and limitations of the present framework and provide a basis for improving its robustness and extending it toward more demanding applications, such as localized heavy-rainfall forecasting and urban micrometeorology.

The remainder of the paper is organized as follows. Section~\ref{sec:method} formulates the problem and describes the BridgeSRDA methodology, training and cycling procedures. Section~\ref{sec:setup} presents the QG testbed and evaluation design. Section~\ref{sec:results} reports the results, and Section~\ref{sec:discussion} discusses their implications. Section~\ref{sec:conclusions} concludes the paper and outlines directions for future work. Additional methodological and experimental details are provided in the appendices.

\section{Problem formulation and methodology}\label{sec:method}

\subsection{Assimilation setting and spatiotemporal windows}

Let $x(t)$ denote the state of a dynamical system. At an assimilation time $t_c$, the observations are given by
\begin{equation}
y_c=\mathcal H_c(x(t_c))+e_c,
\label{eq:observations}
\end{equation}
where $\mathcal H_c$ maps the state to observation space and $e_c$ represents observation error. The task of data assimilation is to combine the observations available up to $t_c$ with a numerical forecast to estimate the current state.

SRDA reduces the cost of this task by advancing a numerical model on a coarse grid and reconstructing the state on a finer grid. In hybrid approaches, learned super-resolution is followed by a conventional ensemble update~\citep{Barthelemy2022}. Alternatively, a neural network can incorporate the observations directly while reconstructing the high-resolution state~\citep{Yasuda2023SRDA,Notsu2025,Ayapilla2026DiffSRDA}. BridgeSRDA follows the latter approach and generates an ensemble of HR analysis windows. Each member's current analysis is downsampled to initialize its next LR forecast.

Each analysis window contains nine frames at physical times
\begin{equation}
t_j=t_c+(j-4)\delta t,\qquad j=0,\ldots,8,\qquad \delta t=0.25.
\label{eq:window}
\end{equation}
The window therefore spans one physical time unit before and after the current assimilation time $t_c$. The LR numerical model supplies forecasts at $t_c-1$, $t_c$ and $t_c+1$. These three forecast fields are spatially lifted to the HR grid and linearly interpolated in physical time to form a nine-frame source window, denoted by $Y_0$. During training, the corresponding HR window provides the target $Y_1$. Observations at $t_c-1$ and $t_c$ provide the additional conditioning information, denoted by $C_{\rm obs}$.

BridgeSRDA transforms the complete source window jointly toward an HR analysis window. The subscripts in $Y_0$ and $Y_1$ refer to the endpoints of this transformation, while $j$ indexes the frames within a window. The two observation representations used in this study are introduced in Section~\ref{sec:observations}.

\subsection{Conditional stochastic-interpolant transport}

\begin{figure}
  \centering
  \includegraphics[width=1.0\textwidth]{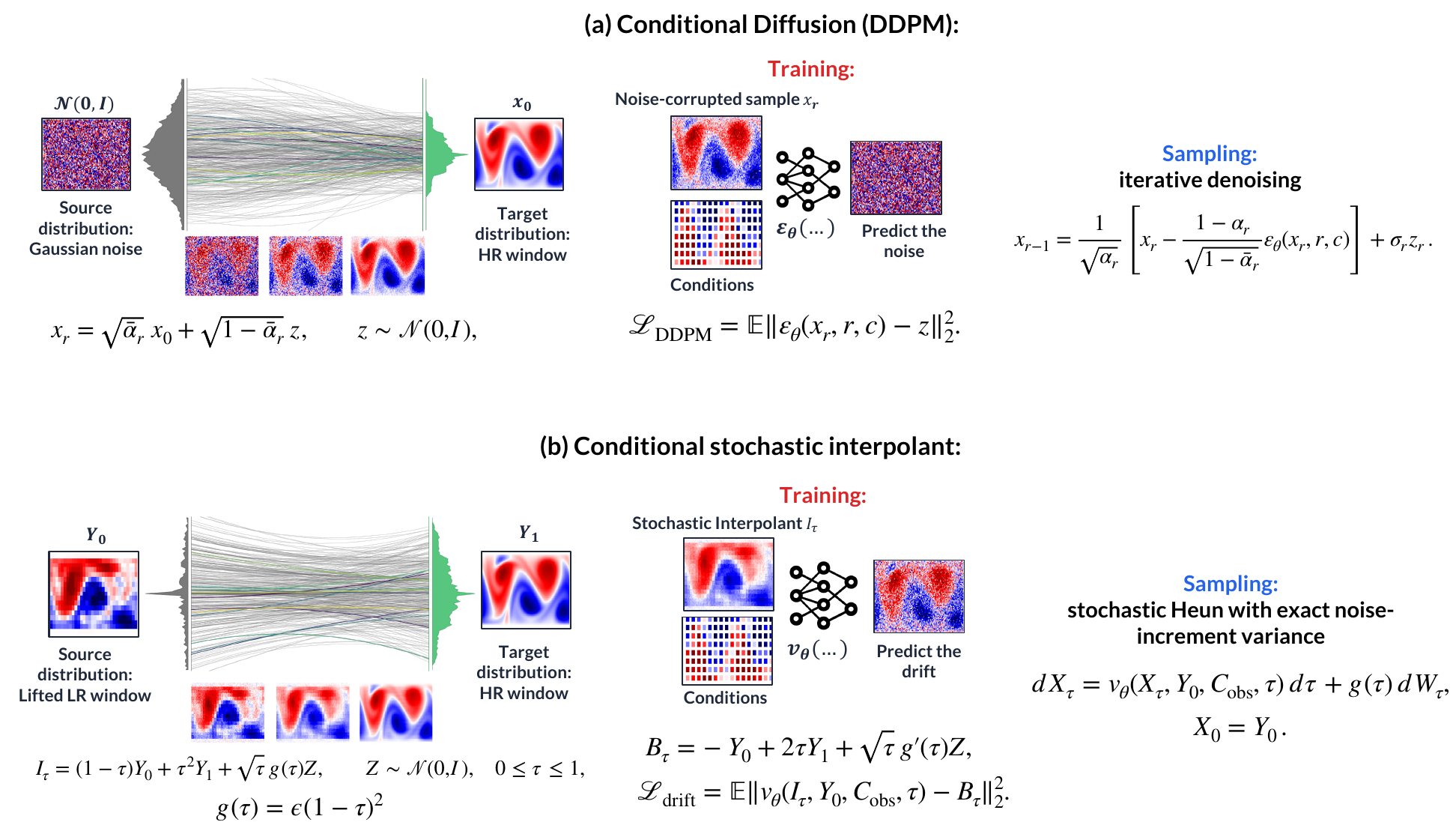} \caption{Conditional generative formulations for SRDA. (a) A DDPM learns to predict noise and generates an HR window by iterative denoising from Gaussian noise. (b) BridgeSRDA learns an observation-conditioned drift and transports a lifted LR window toward the HR distribution using stochastic Heun integration. The curves and projected distributions are conceptual illustrations. DDPM generation decreases the corruption index $r$, whereas interpolant transport increases $\tau$. The clean DDPM endpoint is $x_0$, and the interpolant endpoints are $Y_0$ and $Y_1$.}
\label{fig:diffusion-vs-SI}
\end{figure}

We use a conditional stochastic interpolant to define a transformation from the lifted LR forecast window $Y_0$ to the HR target window $Y_1$. The transformation is indexed by an artificial transport time $\tau\in[0,1]$, which is distinct from the physical time of the flow. As illustrated in Figure~\ref{fig:diffusion-vs-SI}, a conditional DDPM starts from Gaussian noise and learns to remove it, whereas BridgeSRDA starts from the forecast-informed source and learns a drift that transports it toward the HR target distribution.

Following the conditional construction of \citet{Chen2024ConditionalInterpolants}, which has also been used for urban super-resolution~\citep{Yasuda2026Bridge}, we define the normalized training state
\begin{equation}
I_\tau=(1-\tau)Y_0+\tau^2Y_1+\sqrt{\tau}\,g(\tau)Z,
\qquad Z\sim\mathcal N(0,\mathbf I),
\label{eq:interpolant}
\end{equation}
where $g(\tau)=\epsilon(1-\tau)^p$ controls the strength of the stochastic term. At the endpoints, $I_0=Y_0$ and $I_1=Y_1$. The supervised drift target is
\begin{equation}
B_\tau=-Y_0+2\tau Y_1+\sqrt{\tau}\,g'(\tau)Z.
\label{eq:drift-target}
\end{equation}
This target follows from a Brownian interpretation of the interpolant. At a fixed transport time, $\sqrt{\tau}Z$ has the distribution of a Wiener process $W_\tau$, and $B_\tau$ is the corresponding It\^o drift. It differs from the ordinary derivative obtained by holding one Gaussian tensor fixed as $\tau$ varies. Appendix~\ref{app:brownian} gives the derivation.

We use a quadratic noise envelope (p2) with $p=2$ and $\epsilon=0.2581988897471611$. This envelope reduces the amount of new noise added near the HR endpoint. The amplitude is chosen so that the total sampling-noise variance matches that of the linear envelope (p1; $p=1$, $\epsilon=0.2$) used in the comparison. Appendix~\ref{app:noise} explains the effect of the envelope on both sampling and drift supervision and presents the validation comparison.

A neural network $v_\theta$ takes the training state $I_\tau$, the source $Y_0$, the observation conditions $C_{\rm obs}$ and the transport time $\tau$ as inputs. It is trained to predict the drift by minimizing
\begin{equation}
\mathcal L_{\rm drift}=\mathbb E\left[
\|v_\theta(I_\tau,Y_0,C_{\rm obs},\tau)-B_\tau\|_2^2\right].
\label{eq:loss}
\end{equation}
At the population optimum, the network estimates the mean of the target drift conditional on its inputs.

To generate an HR window, we start from $X_0=Y_0$ and use the learned drift in the stochastic differential equation
\begin{equation}
dX_\tau=v_\theta(X_\tau,Y_0,C_{\rm obs},\tau)d\tau+g(\tau)dW_\tau,
\qquad X_0=Y_0.
\label{eq:sampling-sde}
\end{equation}
The drift moves the state toward the HR target distribution, while the Wiener term introduces stochastic variability. Under ideal regression and suitable regularity assumptions, the continuous process has the conditional marginals prescribed by the training construction. In practice, finite training, numerical integration and differences between training and deployment inputs affect this correspondence.

This construction is related to Schr\"odinger bridges, which minimize path-space relative entropy with respect to a reference process subject to endpoint constraints. BridgeSRDA uses a prescribed stochastic interpolant and samples with the directly learned drift and its matching diffusion coefficient. \citet{Chen2024ConditionalInterpolants} also describe a F\"ollmer construction that adjusts both the drift and diffusion and, under their stated assumptions, has a Schr\"odinger-bridge interpretation relative to a specified reference process. The present sampler does not use that adjustment, and we have not established Schr\"odinger-bridge optimality for it. We therefore describe BridgeSRDA as conditional stochastic-interpolant transport. Its contribution is the adaptation of this construction to observation-conditioned trajectory analysis and sequential forecasting with a numerical model.

\subsection{Training and sequential assimilation with BridgeSRDA}

\subsubsection{Offline training}

\begin{figure}
  \centering
  \includegraphics[width=1.0\textwidth]{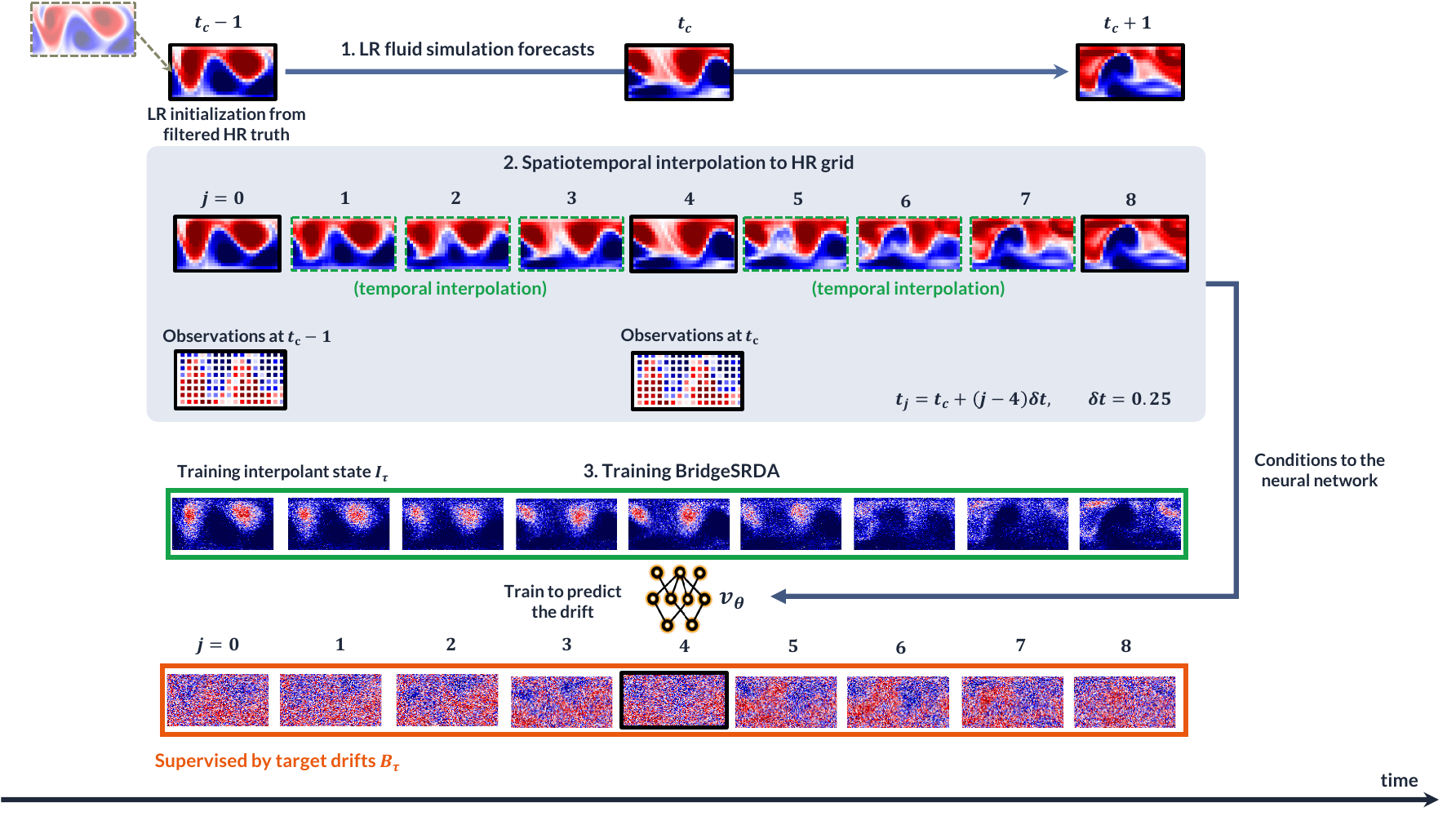} \caption{Construction of training windows for BridgeSRDA. LR forecasts initialized from filtered HR truth provide three fields for a lifted nine-frame source. The source and past/current observations condition a neural network that is trained against the drift target at a sampled transport time. The frames span physical time, while all frames of an interpolant share the same transport time. The temporal evolution of the snapshots is exaggerated for visual clarity.}
\label{fig:bridgesrda-training-overview}
\end{figure}

Figure~\ref{fig:bridgesrda-training-overview} summarizes how the training windows are constructed. HR training trajectories provide target windows and synthetic observations. For each window, the LR numerical model is initialized from filtered HR truth and advanced to obtain the three coarse forecast fields. Spatial lifting and temporal interpolation then form the source window $Y_0$. The observations available at the past and current assimilation times form $C_{\rm obs}$.

For each paired source and target window, a transport time $\tau$ and a Gaussian tensor $Z$ determine the training state $I_\tau$ and target drift $B_\tau$ through Equations~\eqref{eq:interpolant} and \eqref{eq:drift-target}. The network receives $\tau$ as an input and learns to predict the drift from the corresponding training state and conditions.

\begin{figure}
  \centering
  \includegraphics[width=1.0\textwidth]{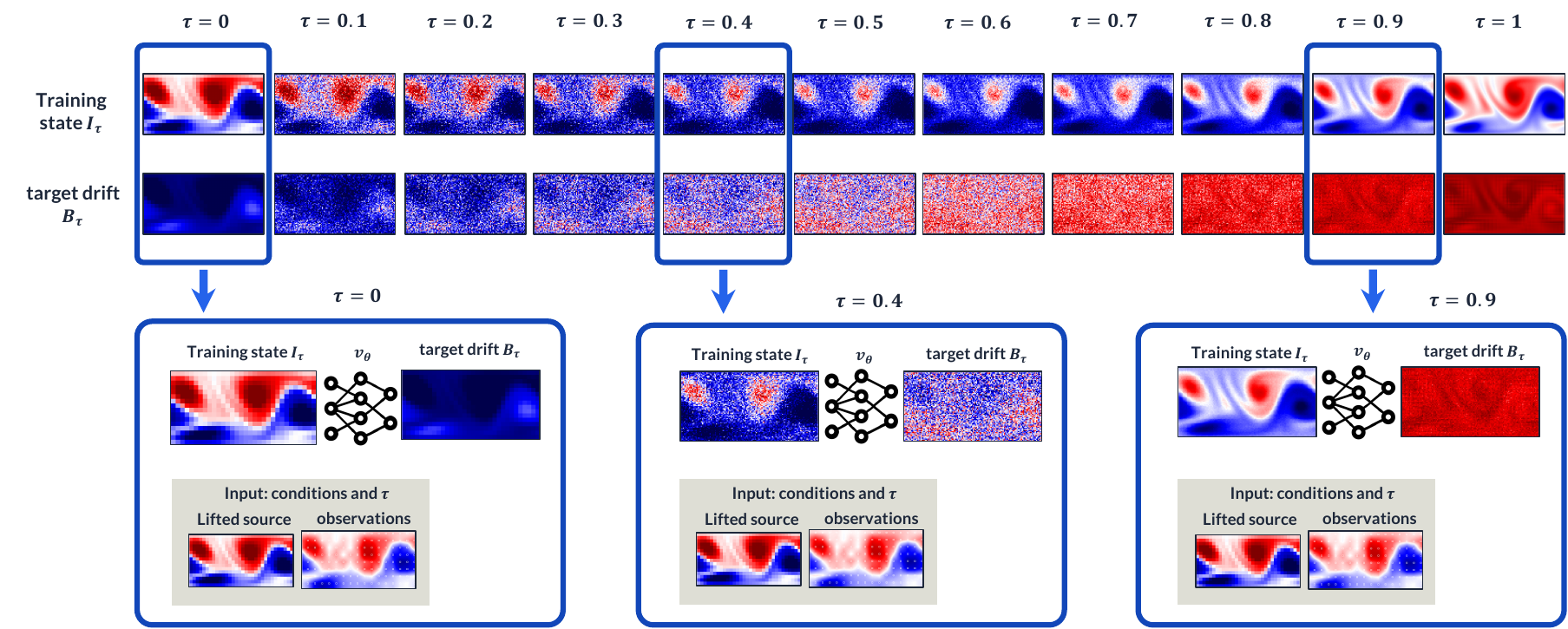} \caption{Drift supervision at selected transport times, illustrated for the middle-layer current frame at physical time 100. The source $Y_0$ and observation conditions remain fixed as the training state $I_\tau$ varies. The network output $v_\theta$ is compared with the target drift $B_\tau$. Training samples transport times across windows rather than presenting the displayed states as a chronological sequence. The panels use illustrative coarse proxies and synthetic observations.}
\label{fig:bridgesrda-training-detail}
\end{figure}

Figure~\ref{fig:bridgesrda-training-detail} illustrates this process for one physical-time frame at several transport times. The displayed states are examples of possible training inputs; they are not supplied to the network as a chronological sequence. Within each effective batch of 32 windows, we include both transport endpoints and sample one time uniformly from each of 30 interior strata. The selected times are randomly assigned to windows and renewed during training, providing coverage across the transport interval.

\subsubsection{Inference-time cycling}

\begin{figure}
  \centering
  \includegraphics[width=1.0\textwidth]{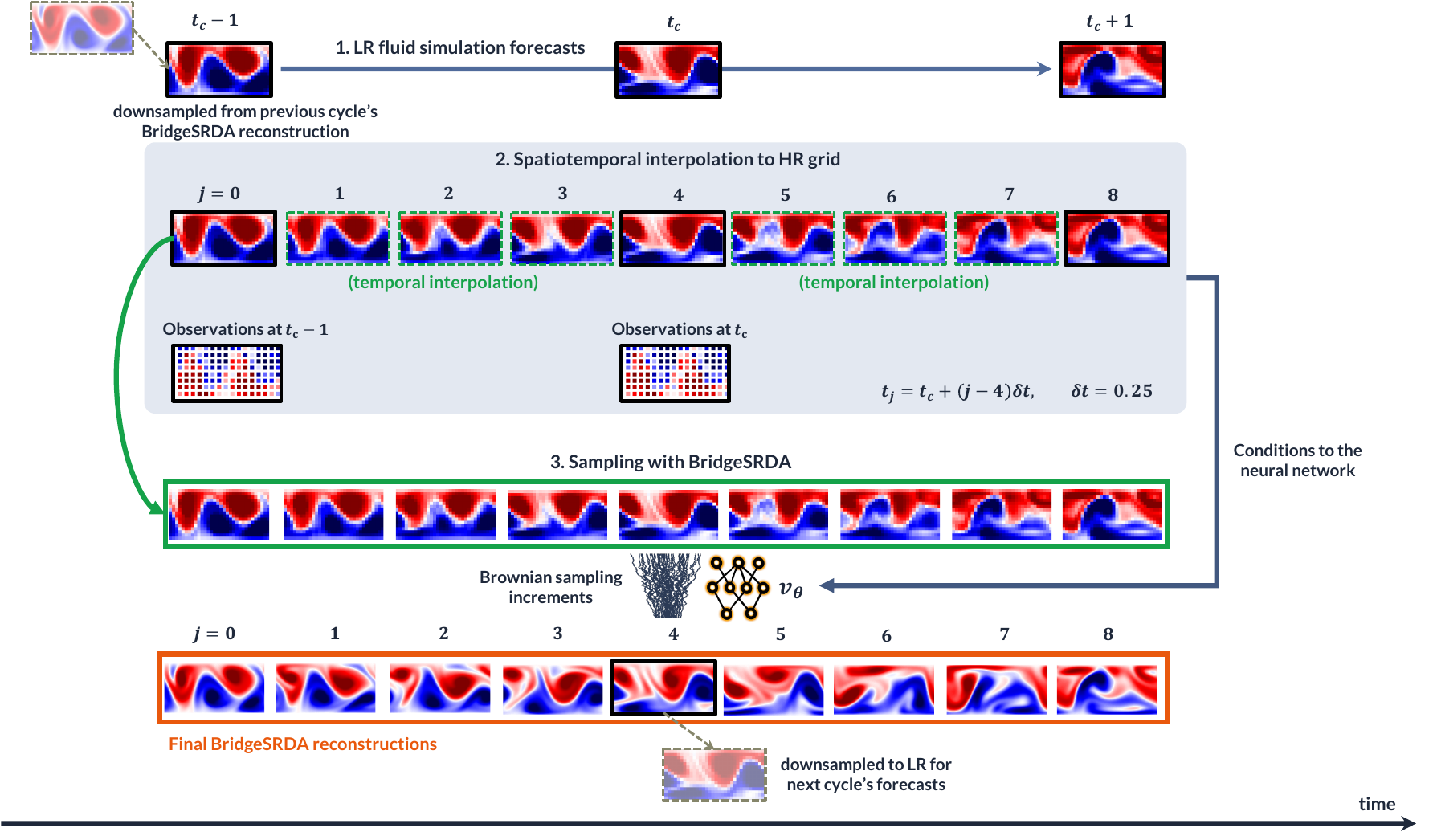} \caption{Inference-time reconstruction and cycling with BridgeSRDA. Each member's LR forecast provides three fields that are lifted and interpolated in time to form a nine-frame source. Observations at $t_c-1$ and $t_c$ provide the conditioning information. The sampled HR window contains retrospective reconstructions at $j=0$--3, a current analysis at $j=4$, and nowcasts at $j=5$--8, corresponding to physical horizons of 0.25, 0.50, 0.75 and 1.00. The nowcasts require no future observations or additional sampling calls. Each member's current analysis is downsampled to initialize its next LR forecast. The temporal evolution of the snapshots is exaggerated for visual clarity; the actual window spans two physical time units.}
\label{fig:bridgesrda-inference-cycling}
\end{figure}

During deployment, the trained model is used within a repeated forecast--analysis cycle, as shown in Figure~\ref{fig:bridgesrda-inference-cycling}. For each ensemble member, the current HR analysis from the previous cycle is downsampled to initialize an LR numerical forecast. The resulting LR forecast window and the available observations condition the stochastic sampler. Independent sampling increments and member-specific forecasts produce an ensemble of HR windows.

The nine frames in each sampled window serve three roles. Frames $j=0$--3 reconstruct the recent past, frame $j=4$ is the current analysis, and frames $j=5$--8 provide short-range nowcasts. Only each member's current analysis is fed back to initialize its next LR forecast. The future frames use the LR forecast extending beyond $t_c$, but they do not use observations from after $t_c$.

The retrospective frames have a smoothing interpretation because the observations available at $t_c$ can inform states before that time. The central frame provides a filtering-type current analysis, while the remaining frames provide nowcasts. This smoothing is limited to the current window; the method does not revisit the complete history of past states.

The main experiments use an assimilation interval of one physical time unit. The LR forecast extends one unit beyond the current assimilation time to provide the final part of the source window. During training, the LR forecasts instead begin from filtered HR truth. The possible effect of this difference between training and deployment is discussed in Section~\ref{sec:discussion}.

\subsubsection{Stochastic predictor--corrector sampling}

\begin{figure}
  \centering
  \includegraphics[width=1.0\textwidth]{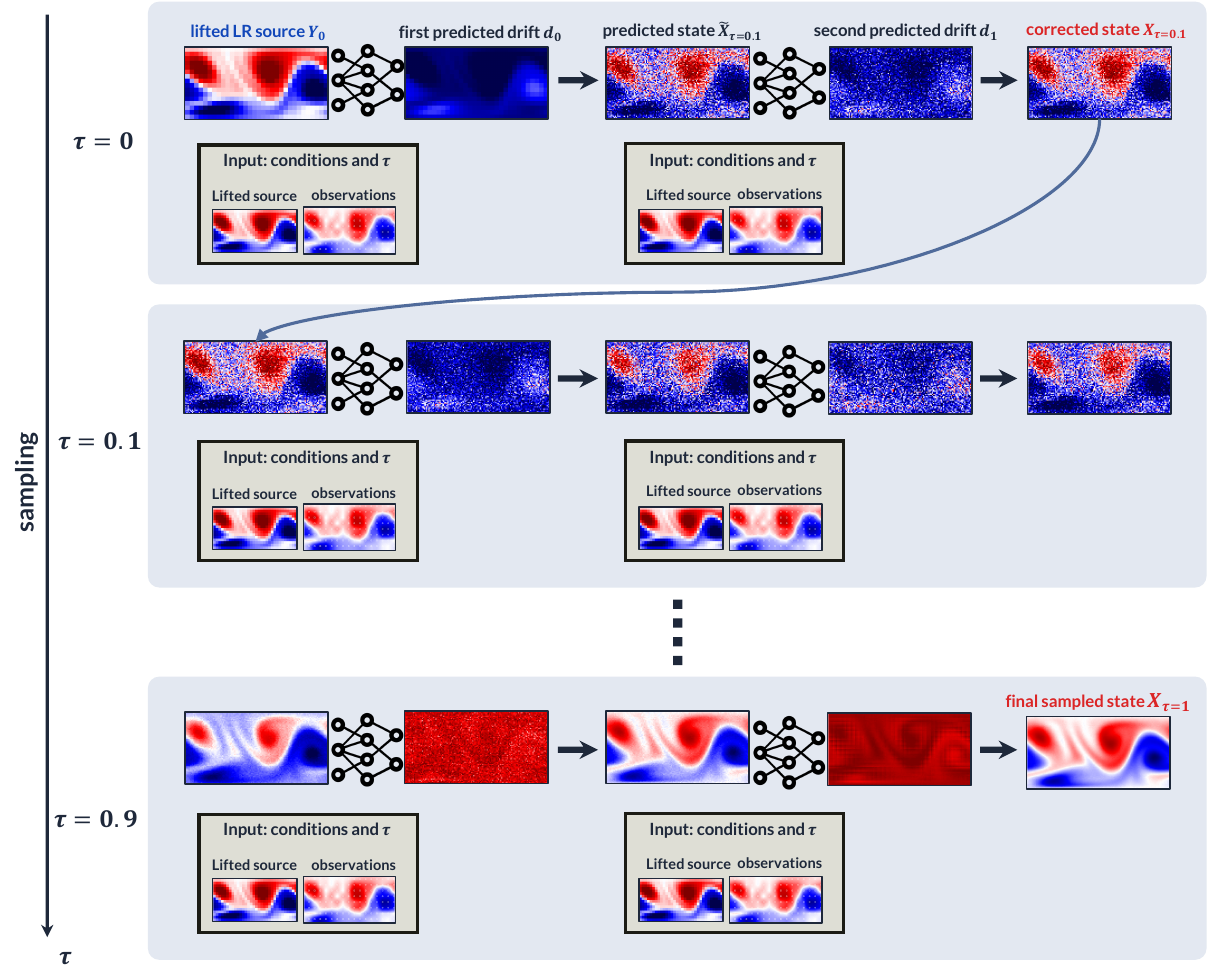} \caption{Stochastic Heun sampling of a nine-frame window, illustrated using the middle-layer current frame at physical time 100. Each interval evaluates the drift at the current state and at a noisy provisional state. The corrected state combines both drift estimates using the same Gaussian increment. The displayed intervals start at transport times 0, 0.1 and 0.9, and the final state is obtained at $\tau=1$. The source and observation conditions remain fixed throughout sampling.}
\label{fig:bridgesrda-heun-sampling}
\end{figure}

We integrate Equation~\eqref{eq:sampling-sde} using stochastic Heun updates. At each interval, a first drift estimate predicts a provisional state. A second drift estimate is evaluated at that state, and the two estimates are combined to correct the update. On a uniform transport-time grid $\tau_k=k/K$ with $K$ intervals, let $X^{[k]}$ denote the numerical state at $\tau_k$. It approximates the continuous state $X_{\tau_k}$. Figure~\ref{fig:bridgesrda-heun-sampling} illustrates these stages, and Algorithm~\ref{alg:heun} gives the complete sampling procedure.

Each interval requires two neural function evaluations (NFE). A sampling budget of NFE20 therefore corresponds to ten Heun intervals, while NFE10 corresponds to five. The Gaussian increment over interval $k$ has variance
\begin{equation}
V_k=\int_{\tau_k}^{\tau_{k+1}}g(s)^2\,ds
=\frac{\epsilon^2}{2p+1}
\left[(1-\tau_k)^{2p+1}-(1-\tau_{k+1})^{2p+1}\right].
\label{eq:increment-variance}
\end{equation}

\begin{algorithm}[tbp]
\caption{Conditional stochastic Heun sampling of one HR window/member}
\label{alg:heun}
\begin{algorithmic}[1]
\Require Drift $v_\theta$, lifted source $Y_0$, observations and masks,
even evaluation budget $\mathrm{NFE}$, member-specific randomness
\State Prepare fixed observation conditions $C_{\rm obs}$ once
\State $p\gets2$, $\epsilon\gets0.2581988897471611$,
$K\gets\mathrm{NFE}/2$, $\Delta\tau\gets1/K$, $X^{[0]}\gets Y_0$
\For{$k=0,\ldots,K-1$}
\State $\tau_k\gets k/K$, $\tau_{k+1}\gets(k+1)/K$
\State $V_k\gets\dfrac{\epsilon^2}{2p+1}
[(1-\tau_k)^{2p+1}-(1-\tau_{k+1})^{2p+1}]$
\State Draw $\xi_k\sim\mathcal N(0,V_k\mathbf I)$
\State $d_k^{(0)}\gets v_\theta(X^{[k]},Y_0,C_{\rm obs},\tau_k)$
\State $\widetilde X^{[k+1]}\gets X^{[k]}+\Delta\tau d_k^{(0)}+\xi_k$
\State $d_k^{(1)}\gets v_\theta(\widetilde X^{[k+1]},Y_0,C_{\rm obs},\tau_{k+1})$
\State $X^{[k+1]}\gets X^{[k]}+\dfrac{\Delta\tau}{2}(d_k^{(0)}+d_k^{(1)})+\xi_k$
\EndFor
\State \Return $X^{[K]}$ at $\tau=1$
\end{algorithmic}
\end{algorithm}

The same Gaussian increment is used in both stages of each interval, and the corrected state is formed from $X^{[k]}$. The final update reaches $\tau=1$. The source and observation conditions remain fixed throughout sampling. The expression for $V_k$ gives the exact variance of the integrated noise over each interval, although the numerical integration of the learned drift remains approximate.

\section{Quasi-geostrophic testbed, training setup and evaluation}\label{sec:setup}

\subsection{Dynamical testbed and observation representations}\label{sec:observations}

We evaluate BridgeSRDA using an observing-system simulation based on a multilayer QG baroclinic-jet benchmark~\citep{Notsu2025}. The model describes the evolution of potential vorticity (PV) in a channel with dimensions $L_x=2\pi$, $L_y=\pi$ and $L_z=0.0125$. The domain is periodic in the zonal direction $x$ and nonperiodic in the meridional direction $y$.

PV anomalies $q$ are related to the streamfunction $\psi$ through horizontal and vertical coupling. The streamfunction also determines the horizontal velocity components $u$ and $v$. In continuous shorthand for the layer-discretized model,
\begin{equation}
q=\nabla_h^2\psi+\mathcal L_z\psi,\qquad
u=-\partial_y\psi,\quad v=\partial_x\psi.
\label{eq:qg-inversion}
\end{equation}
The PV anomalies evolve according to
\begin{equation}
\partial_tq=-[(u+U)\partial_xq+v\partial_yq+vQ_y]+\mathcal D(q)+f_q.
\label{eq:qg-dynamics}
\end{equation}
Here, $\mathcal L_z$ describes vertical PV coupling, $U$ is the horizontally averaged zonal velocity, and $Q_y=\beta-\mathcal L_zU$ includes the planetary PV gradient $\beta$. The operator $\mathcal D$ represents spectral hyperdiffusion. The meridionally varying jet is represented through $q$ and its inverted streamfunction, and no external forcing is applied ($f_q=0$). The physical velocity $v$ is distinct from the neural transport drift $v_\theta$.

\begin{figure}
  \centering
  \includegraphics[width=1.0\textwidth]{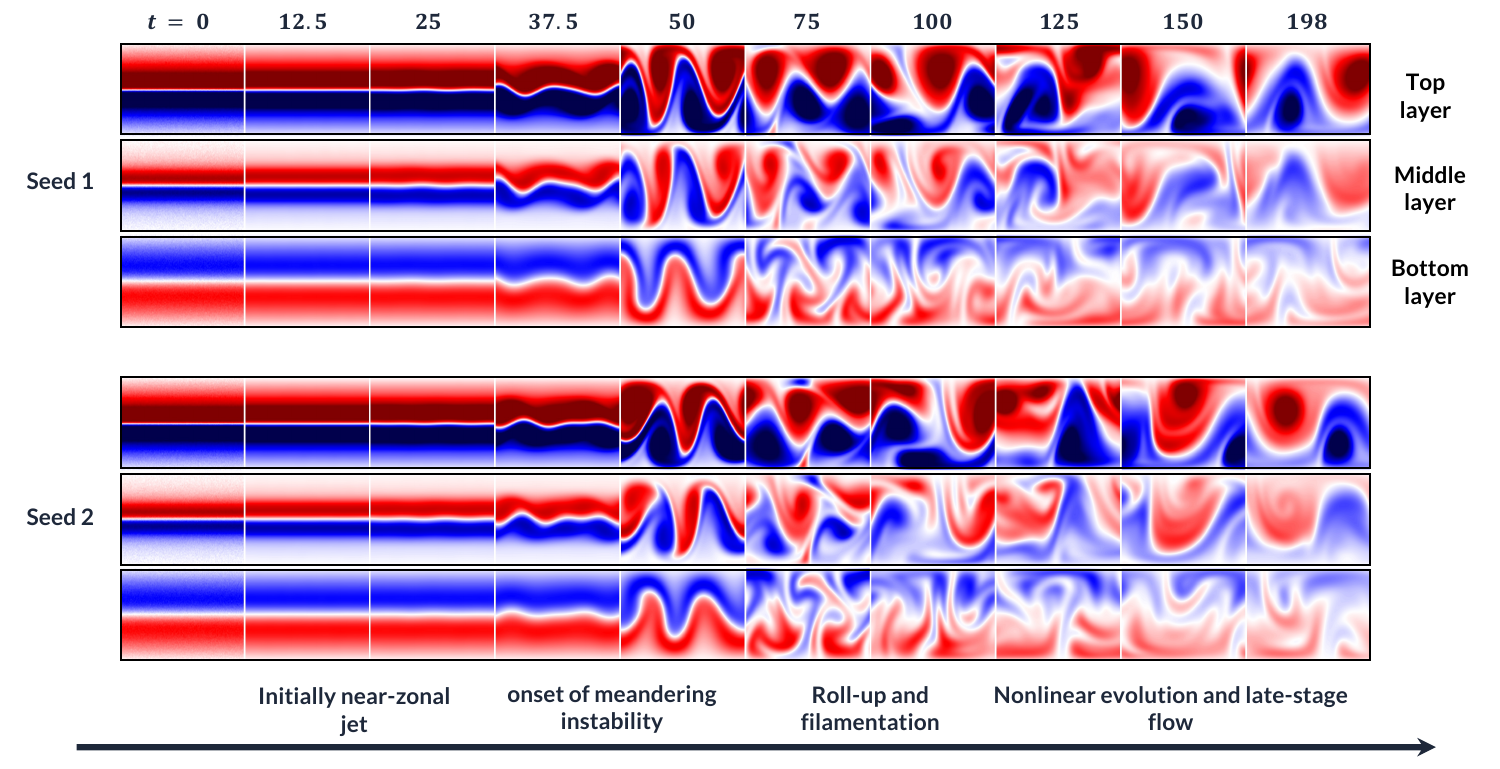} \caption{Evolution of PV in two independently initialized trajectories of the multilayer baroclinic-jet system. For each trajectory, the rows show the top, middle and bottom layers and the columns show physical time. The initially near-zonal jet develops meanders, rolls up and forms filaments as the flow becomes nonlinear.}
\label{fig:baroclinic-evolution}
\end{figure}

Figure~\ref{fig:baroclinic-evolution} shows how an initially near-zonal jet develops into a nonlinear flow through meandering, roll-up and filamentation. Different initial perturbations produce distinct trajectories. The multilayer system therefore provides a controlled setting in which to assess both observation assimilation and reconstruction across the vertical layers.

We use LR forecasts, HR analyses and an ultra-high-resolution (UHR) nature run. Their native array shapes are $4\times17\times32$, $16\times65\times128$ and $32\times129\times256$, respectively, ordered as $z/y/x$. Neural preprocessing removes the last meridional row, giving effective LR and HR shapes of $4\times16\times32$ and $16\times64\times128$. The HR grid is therefore four times finer than the LR grid in each spatial direction. To construct the source window, we lift the LR forecasts using nearest-exact spatial interpolation and then interpolate linearly in time between the three forecast anchors. Appendix~\ref{app:implementation} provides the numerical parameters and details of the reference conversion.

The test cases differ from the training data in both resolution and jet width. Training targets are HR trajectories generated with a meridional jet-width parameter of 0.4. The UHR nature run instead uses a width of 0.36, producing an initially narrower jet. We refer to this test configuration as the narrow jet. Its finer grid provides the synthetic observations and the reference solution. The reference is converted to the HR grid for evaluation. All methods are tested on the narrow-jet trajectories.

The synthetic observations measure PV on a regular horizontal lattice at every vertical layer. Independent Gaussian noise with a standard deviation of 0.022 in physical PV units is added to the measurements. During training, the lattice spacing is eight HR grid points in each horizontal direction, while its origin varies.

\begin{figure}
  \centering
  \includegraphics[width=0.7\textwidth]{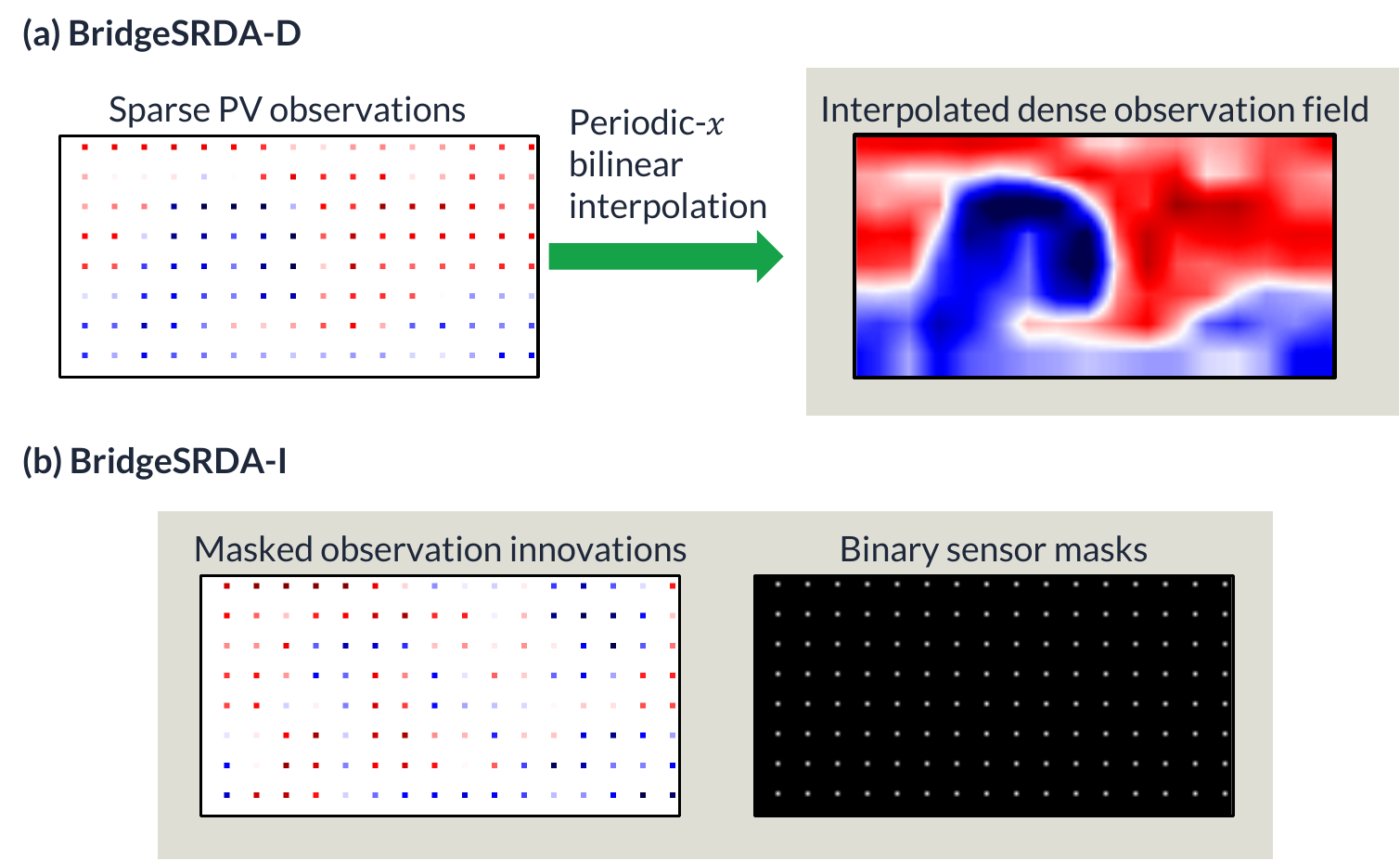} \caption{Observation representations used by BridgeSRDA-D and BridgeSRDA-I. BridgeSRDA-D interpolates sparse PV measurements to dense fields. BridgeSRDA-I uses masked observation-minus-source innovations and binary sensor masks. The panels illustrate the preprocessing for a horizontal slice at an observation time. Interpolated values are not additional measurements. Both models also receive the lifted source and transport-time embedding.}
\label{fig:observation-preprocessing}
\end{figure}

We consider two ways of supplying these observations to BridgeSRDA. \textbf{BridgeSRDA-D} uses interpolated dense observation fields, whereas \textbf{BridgeSRDA-I} uses sparse innovations together with sensor masks. Figure~\ref{fig:observation-preprocessing} illustrates the two representations.

For BridgeSRDA-D, the past and current observations are interpolated independently on each layer to produce two dense HR fields. The bilinear interpolation wraps periodically in $x$ and extends the nearest values to the edges in the nonperiodic $y$ direction. Although the interpolated fields preserve the measured values, the values between sensors do not constitute additional independent observations.

BridgeSRDA-I retains the sparse measurement locations. At each observation time, the normalized measurements are placed on the HR grid to form $O^{[j]}$, and a binary mask $\mathsf M^{[j]}$ identifies their locations. The observation-minus-source innovations are
\begin{equation}
\eta^{[j]}=\mathsf M^{[j]}\odot(O^{[j]}-Y_0^{[j]}),\qquad j\in\{0,4,8\}.
\label{eq:innovations}
\end{equation}
The masks identify where observations are available, distinguishing unobserved locations from measured locations with a zero innovation. Because observations are unavailable at the future endpoint, the future innovation and mask are zero. Both configurations also receive the nine-frame source window and the current transport state. We use BridgeSRDA-D as the primary configuration and compare it with BridgeSRDA-I.

\subsection{Training setup}

Training examples pair LR forecast windows with HR target trajectories. The LR forecasts are initialized from filtered HR truth, as described in Section~\ref{sec:method}.

The training data consist of 23,880 overlapping windows drawn from six initial-condition families. Each family contains 20 member trajectories, with 199 windows per trajectory. Separate families are used for validation and testing so that overlapping windows from the same family do not appear in different subsets. PV values are normalized using bounds of $-2.88$ and $2.92$, with clipping applied before the fields are passed to the network. We also use SR-mixup to augment similar LR trajectories with weights drawn from a Beta(2,2) distribution, while retaining the associated target and observation pairing~\citep{Yasuda2023SRDA}.

Both BridgeSRDA configurations use a residual three-dimensional U-Net with widths of 32/64/128/256/256, attention at the deepest level and a transport-time embedding. The spatial convolution axes are $z/y/x$, while the physical-time frames are represented as input and output channels.

We train the networks using adaptive moment estimation with decoupled weight decay (AdamW), a learning rate of $10^{-4}$, zero weight decay and gradient clipping at one. We also maintain an exponential moving average (EMA) of the network weights with decay 0.999. Each epoch presents 1,920 training samples and 96 validation samples. Each effective batch of 32 includes the two transport endpoints and one randomly sampled time from each of 30 interior strata.

We monitor training and validation losses together with validation reconstructions. The main cycling evaluations use the EMA weights. Additional network and sampling details are given in Appendix~\ref{app:implementation}.

\subsection{Sampling and evaluation}

We evaluate all methods on ten independently initialized narrow-jet trajectories. For each trajectory, the main comparison covers 199 assimilation times, $t_c=1,\ldots,199$, with one assimilation per physical time unit.

The primary generative configuration uses 20 neural function evaluations and 30 ensemble members, denoted NFE20/M30. To examine the effects of sampling budget and ensemble size, we also evaluate NFE20/M10 and NFE10 with either 30 or ten members. These comparisons use the same trained model weights.

In equations, $N_{\rm ens}$ denotes the number of ensemble members. In figures and tables, M10 and M30 refer to ensembles of ten and 30 members, respectively. The symbol $\mathsf M$, by contrast, denotes an observation mask.

\subsection{Reconstruction and ensemble metrics}\label{sec:metrics}

We evaluate the primary scores on the HR grid against the converted UHR nature run. For an ensemble-mean prediction $\bar q$ and reference $q^\star$, the mean absolute error ratio (MAER) and root-mean-square error (RMSE) are
\begin{equation}
\mathrm{MAER}=\frac{\langle|\bar q-q^\star|\rangle}{\langle|q^\star|\rangle+\delta},
\qquad \mathrm{RMSE}=\sqrt{\langle(\bar q-q^\star)^2\rangle},
\label{eq:point-metrics}
\end{equation}
where angle brackets denote spatial averages and $\delta$ is a small denominator safeguard.

To assess spatial structure, we use a structural loss defined as one minus the mean structural similarity (MSSIM), calculated on normalized and clipped fields. We also calculate RMSE for horizontal Laplacian PV to assess differences in small-scale curvature. The Laplacian uses a unit-grid finite-difference stencil with periodic treatment in $x$ and nearest-edge extension in $y$. It does not include physical grid-spacing factors.

We assess ensemble uncertainty using spread, the continuous ranked probability score (CRPS), interval coverage and rank histograms. Ensemble spread is the spatial root-mean-square (RMS) of the sample standard deviation, calculated with variance denominator $N_{\rm ens}-1$. For ensemble members $q^{(m)}$, the empirical CRPS at a grid point is
\begin{equation}
\mathrm{CRPS}=\frac{1}{N_{\rm ens}}\sum_m|q^{(m)}-q^\star|
-\frac{1}{2N_{\rm ens}^2}\sum_{m,n}|q^{(m)}-q^{(n)}|.
\label{eq:crps}
\end{equation}
Coverage is the fraction of reference values between the ensemble's 5th and 95th percentiles. Rank histograms provide a further assessment of how the reference falls within the ensemble distribution~\citep{Hamill2001Ranks}.

Scores are averaged spatially and then over assimilation times and independent trajectories. Spread/skill is calculated by averaging the per-cycle ratios of spread to RMSE. Grid points and overlapping times are correlated and are not treated as independent statistical replicates. Appendix~\ref{app:implementation} describes the boundary treatment and structural-metric implementation.

\subsection{Comparison methods}\label{sec:comparison-methods}

Our main probabilistic reference is the local ensemble transform Kalman filter (LETKF)~\citep{Hunt2007}, with numerical forecasts and assimilation performed on the HR grid. We denote this method HR-LETKF and primarily use 30 ensemble members. We also evaluate HR-LETKF with ten members to examine its sensitivity to ensemble size. LR-LETKF instead performs assimilation on the coarse grid, after which its output is lifted to the HR grid for evaluation.

As a deterministic neural reference, we reproduce four-dimensional SRDA (4D-SRDA;~\citealp{Notsu2025}). Its multi-axis vision transformer (MaxViT) network has 383,725,067 parameters, compared with 90,832,745 for BridgeSRDA-D. BridgeSRDA-D therefore uses approximately 76.3\% fewer parameters. Parameter count describes the size of the networks, whereas deployment cost also depends on the number of neural evaluations and ensemble members.

We also compare BridgeSRDA with SR-LETKF, a hybrid that separates learned SR from the assimilation update. Its independently trained, observation-free SR model uses the same residual U-Net backbone, quadratic source-to-target interpolant, p2 noise envelope and stochastic Heun construction as BridgeSRDA. Unlike the direct assimilation models, it receives no observation channels and is trained only to reconstruct HR forecast windows from LR forecasts.

At inference, the SR model generates HR forecast members using NFE20. LETKF then combines these members with the current observations, and each updated member is downsampled to initialize the next LR forecast. This comparison allows us to examine learned forecast reconstruction followed by a conventional assimilation update alongside BridgeSRDA's direct observation-conditioned transport.

The HR-LETKF configuration uses $6\times6$ horizontal analysis blocks, inflation of 1.2, an observation-selection radius of 6.0 and a localization-decay radius of 5.7 in domain coordinates. Its configured observation-covariance diagonal is 0.022, while the variance used to generate the synthetic observation errors is $0.022^2=0.000484$. The filter therefore assumes a larger observation variance than that used to generate the measurements. Further settings are given in Appendix~\ref{app:implementation}.

Table~\ref{tab:method-labels} summarizes the method and reference labels used throughout the results.

\begin{table}[!htbp]\centering\small
\caption{Methods and reference fields used in the evaluation. M10 and M30 denote ensembles of ten and 30 members, respectively.}
\label{tab:method-labels}
\begin{adjustbox}{max width=\linewidth}
\begin{tabular}{ll}
\toprule
Label & Description \\
\midrule
BridgeSRDA-D & Direct probabilistic SRDA with dense observation fields. \\
BridgeSRDA-I & Direct probabilistic SRDA with masked innovations and sensor masks. \\
HR-LETKF & LETKF with numerical forecasts and assimilation on the HR grid. \\
LR-LETKF & LETKF on the LR grid, with output lifted to HR for evaluation. \\
SR-LETKF & Observation-free learned SR followed by an HR LETKF update. \\
4D-SRDA & Deterministic neural spatiotemporal SRDA baseline of \cite{Notsu2025}. \\
LR free run & Unassimilated LR forecast, interpolated to the HR grid. \\
HR truth & UHR nature run converted to the HR verification grid. \\
\bottomrule
\end{tabular}
\end{adjustbox}
\end{table}

We measure deployment time per analysis on an NVIDIA RTX A6000 graphics processing unit (GPU). The reported time includes numerical forecasting and the assimilation update or neural sampling, but excludes offline training. The main comparison uses 32-bit floating-point (FP32) network tensors for BridgeSRDA and 64-bit floating-point (FP64) numerical evolution and filtering for HR-LETKF. GPU occupancy and execution conditions can differ between the recorded runs. The resulting times characterize these implementations and conditions, rather than the maximum achievable performance of either method.

We additionally investigate faster execution and smaller BridgeSRDA ensembles on the same hardware. This experiment follows complete recursive cycling on one trajectory and examines forecasting and update costs alongside reconstruction and uncertainty. It retains the usual observation-refresh rate and is reported separately in Appendix~\ref{app:execution-efficiency}. Table~\ref{tab:execution-settings} summarizes the execution settings for the main comparison and this additional experiment.

\begin{table}[!htbp]\centering\small
\caption{Execution settings for the main comparison and the additional efficiency experiment on an NVIDIA RTX A6000 GPU. The efficiency experiment uses one trajectory and changes execution and, for BridgeSRDA, ensemble size without retraining. Its results are reported separately from the main ten-trajectory comparison.}
\label{tab:execution-settings}
\begin{adjustbox}{max width=\linewidth}
\begin{tabular}{lll}
\toprule
Method & Main comparison & Separate efficiency experiment \\
\midrule
BridgeSRDA-D & FP32 tensors, member-wise sampling & Batched, compiled mixed-precision sampling \\
HR-LETKF & FP64 forecast and reference update & FP64 forecast and batched local update \\
\bottomrule
\end{tabular}
\end{adjustbox}
\end{table}

\section{Results}\label{sec:results}

\subsection{Reconstruction accuracy}\label{sec:reconstruction-accuracy}

\begin{figure}
  \centering
  \includegraphics[width=1.0\textwidth]{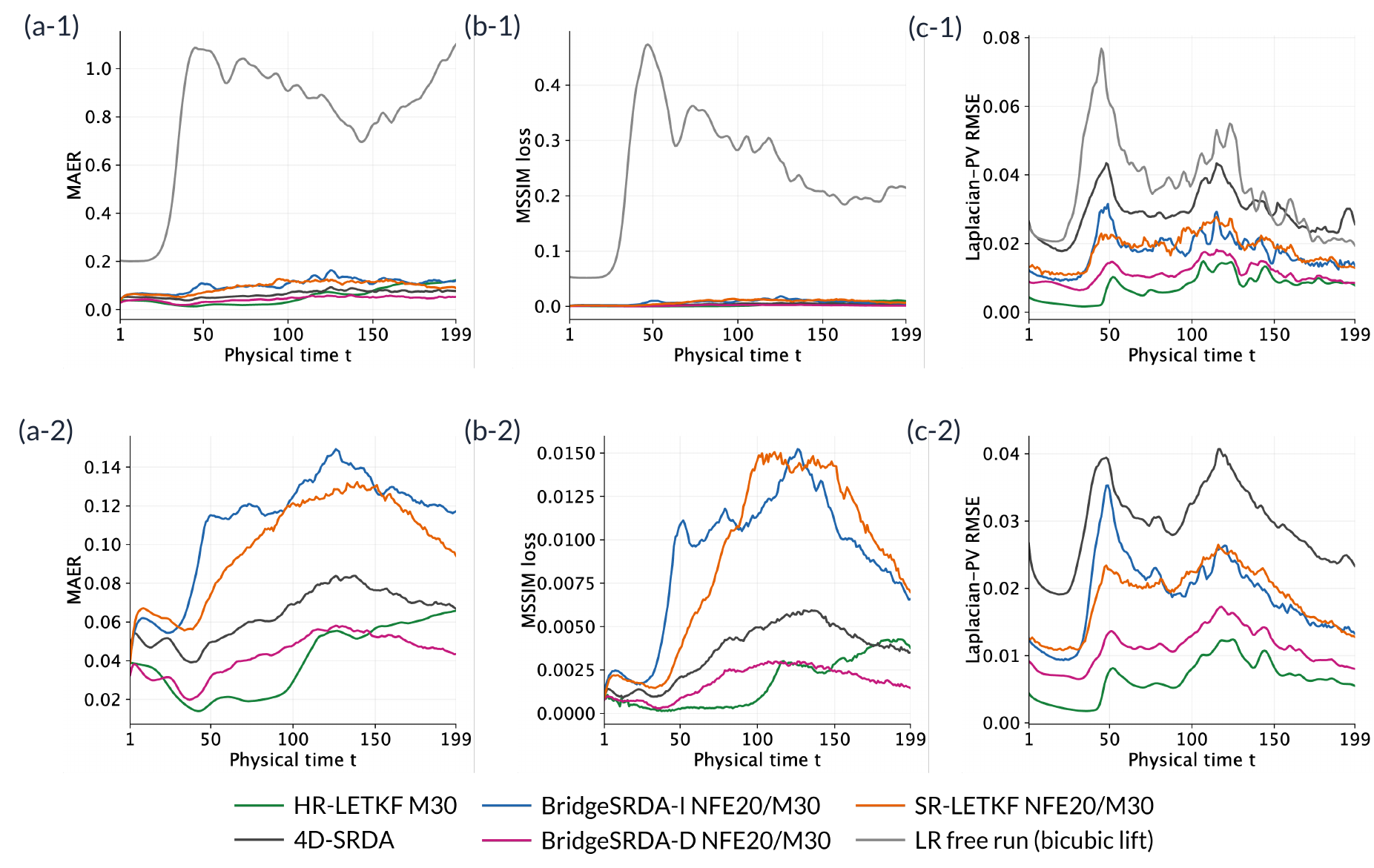} \caption{HR-grid reconstruction errors over physical analysis time $t$. The top row shows results for one narrow-jet trajectory, including an LR free run initialized from the same initial condition and evolved without subsequent observations. The bottom row shows errors averaged over ten independently initialized narrow-jet trajectories. Columns show (a-1, a-2) MAER, (b-1, b-2) MSSIM loss and (c-1, c-2) horizontal Laplacian-PV RMSE. The free-run fields are lifted using bicubic horizontal and linear vertical interpolation. BridgeSRDA-D, BridgeSRDA-I and SR-LETKF use NFE20/M30, and HR-LETKF uses 30 members. Lower values indicate better reconstruction. The Laplacian scores use the unit-grid horizontal stencil defined in Section~\ref{sec:metrics}.}
\label{fig:02-point-skill}
\end{figure}

We first compare the accuracy of the reconstructed HR analyses at the observation spacing used during training. Figure~\ref{fig:02-point-skill}(a-2)--(c-2) shows the evolution of the errors averaged over the ten narrow-jet trajectories, and Table~\ref{tab:canonical-point-scores} summarizes the average scores and recorded costs.

BridgeSRDA-D achieves an MAER of 0.04319, close to 0.03951 for HR-LETKF M30 and lower than 0.06399 for the deterministic 4D-SRDA model. This is an approximately 32.5\% reduction relative to 4D-SRDA, even though BridgeSRDA-D uses about one quarter as many network parameters and generates an ensemble of HR trajectories.

The structural loss shows a similar comparison: BridgeSRDA-D approaches HR-LETKF and improves on the deterministic model. The difference between BridgeSRDA-D and HR-LETKF is more apparent in Laplacian-PV RMSE. Although BridgeSRDA-D substantially improves on 4D-SRDA, HR-LETKF reconstructs fine-grid curvature more accurately. We examine these differences in the flow fields below and assess the quality of the sampled uncertainty in Section~\ref{sec:results-uq}.

\begin{table}[!htbp]\centering\small
\caption{HR-grid reconstruction scores and recorded cost at the observation spacing used during training. Methods are ordered by MAER. Learned ensemble configurations use NFE20/M30, whereas 4D-SRDA produces a single deterministic window. Scores are averaged over ten independently initialized trajectories and 199 analyses per trajectory. The 4D-SRDA time is its total cycling wall time divided by 199 analyses. Runtimes exclude offline training and were not measured under matched GPU occupancy. The structural metric uses the original structural-similarity implementation.}
\label{tab:canonical-point-scores}
\begin{adjustbox}{max width=\linewidth}
\begin{tabular}{lrrrr}
\toprule
Method & MAER & MSSIM loss & Laplacian RMSE & s/analysis \\
\midrule
HR-LETKF M30 & 0.03951 & 0.001719 & 0.00646 & 30.64 \\
BridgeSRDA-D & 0.04319 & 0.001832 & 0.01107 & 16.97 \\
HR-LETKF M10 & 0.04747 & 0.002586 & 0.00678 & 15.83 \\
4D-SRDA & 0.06399 & 0.003605 & 0.02951 & 0.58 \\
SR-LETKF & 0.09991 & 0.008871 & 0.01856 & 34.89 \\
BridgeSRDA-I & 0.11167 & 0.009063 & 0.01891 & 20.33 \\
LR-LETKF M30 & 0.27742 & 0.068061 & 0.03676 & 0.58 \\
\bottomrule
\end{tabular}
\end{adjustbox}
\end{table}

BridgeSRDA-I and SR-LETKF have larger MAER values of 0.11167 and 0.09991, respectively. SR-LETKF nevertheless improves substantially on LR-LETKF, whose MAER is 0.27742. SR-LETKF's recorded cost is approximately 34.9 s per analysis, compared with 16.97 s for BridgeSRDA-D.

In these experiments, directly conditioning the learned reconstruction on observations with BridgeSRDA-D performs better than using an observation-free SR model followed by a LETKF update. The direct approach allows the neural model to use the measurements while generating the fine-scale flow. As described in Section~\ref{sec:observations}, the two BridgeSRDA configurations differ in how they use observations: BridgeSRDA-D receives interpolated observation fields, whereas BridgeSRDA-I receives sparse innovations and sensor masks. We compare their ensemble uncertainty in Section~\ref{sec:results-uq}.

Figure~\ref{fig:02-point-skill}(a-1)--(c-1) includes an LR free run for one test trajectory. The free run starts from the same initial condition but receives no subsequent observations. It departs substantially from the reference, illustrating the importance of repeated assimilation as the flow evolves.

The free run separates less dramatically from the assimilating methods in Laplacian error than in MAER. The two metrics emphasize different aspects of the solution: PV error reflects differences in the flow itself, while the Laplacian emphasizes spatial curvature. Bicubic lifting also smooths the LR field. A comparatively moderate Laplacian error therefore does not necessarily imply an accurate large-scale flow.

\subsection{Reconstructed flow fields and fine-scale structure}

\begin{figure}
  \centering
  \includegraphics[width=1.0\textwidth]{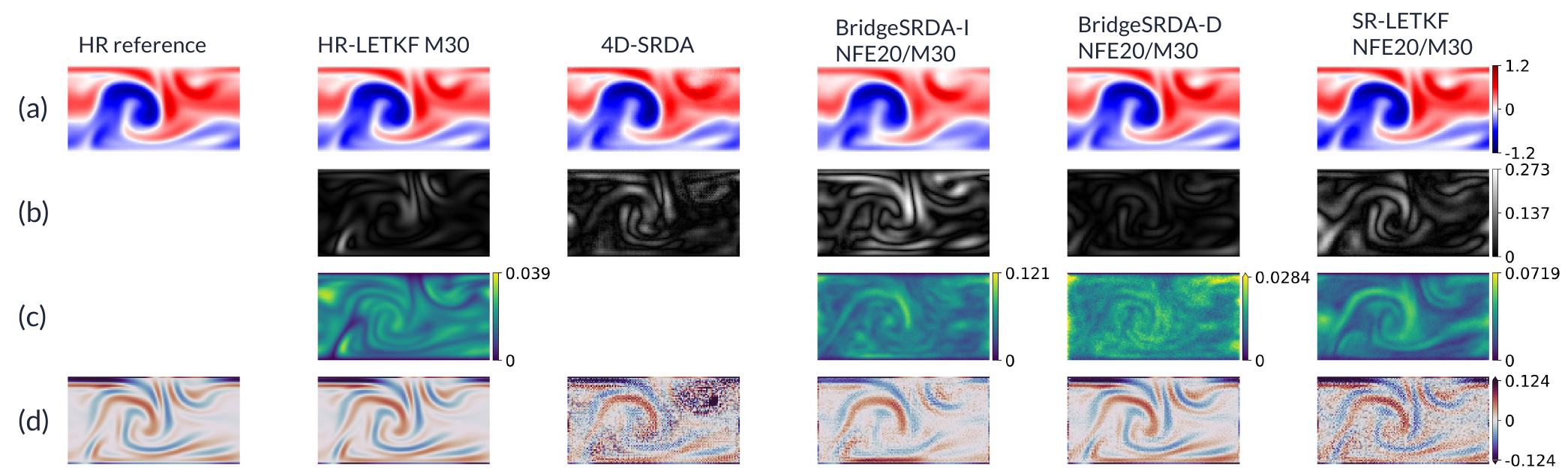} \caption{Analyses at the training observation spacing and physical time 125 for the middle layer, zero-based index 8, of one test trajectory. Columns show the HR reference, HR-LETKF M30, 4D-SRDA, BridgeSRDA-I, BridgeSRDA-D and SR-LETKF. Rows show (a) PV, (b) absolute PV error, (c) ensemble spread and (d) unit-grid horizontal Laplacian PV. The learned ensembles use NFE20/M30. PV, error and Laplacian ranges are shared across methods within the layer. Spread uses method-specific ranges, fixed across the selected times for this layer, so colors do not indicate equal spread magnitudes across methods. Colorbar extensions indicate saturation. Blank cells denote quantities not defined for the reference or deterministic prediction.}
\label{fig:analysis-index0500-middle}
\end{figure}

\begin{figure}
  \centering
  \includegraphics[width=1.0\textwidth]{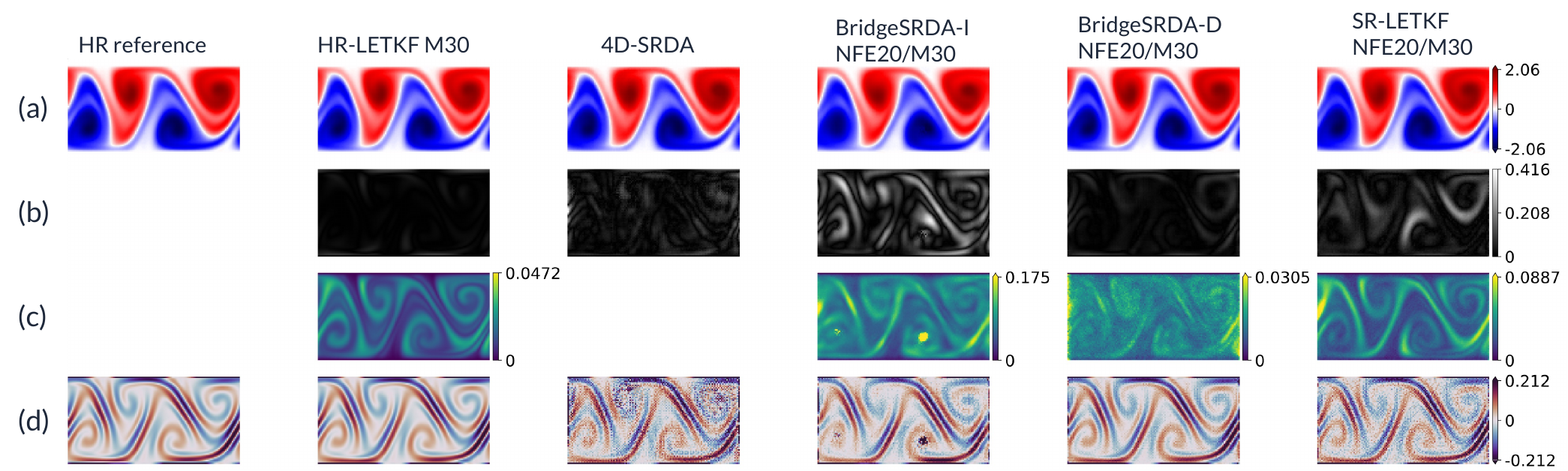} \caption{Analyses at physical time 50 for the top layer, zero-based index 15, of one test trajectory. Columns show the methods and rows show (a) PV, (b) absolute PV error, (c) ensemble spread and (d) unit-grid horizontal Laplacian PV, as in Figure~\ref{fig:analysis-index0500-middle}. Spread uses method-specific ranges, with a colorbar beside each snapshot. The Laplacian range is shared across methods and selected times for this layer. Colorbar extensions indicate saturation.}
\label{fig:analysis-index0200-top}
\end{figure}

\begin{figure}
  \centering
  \includegraphics[width=1.0\textwidth]{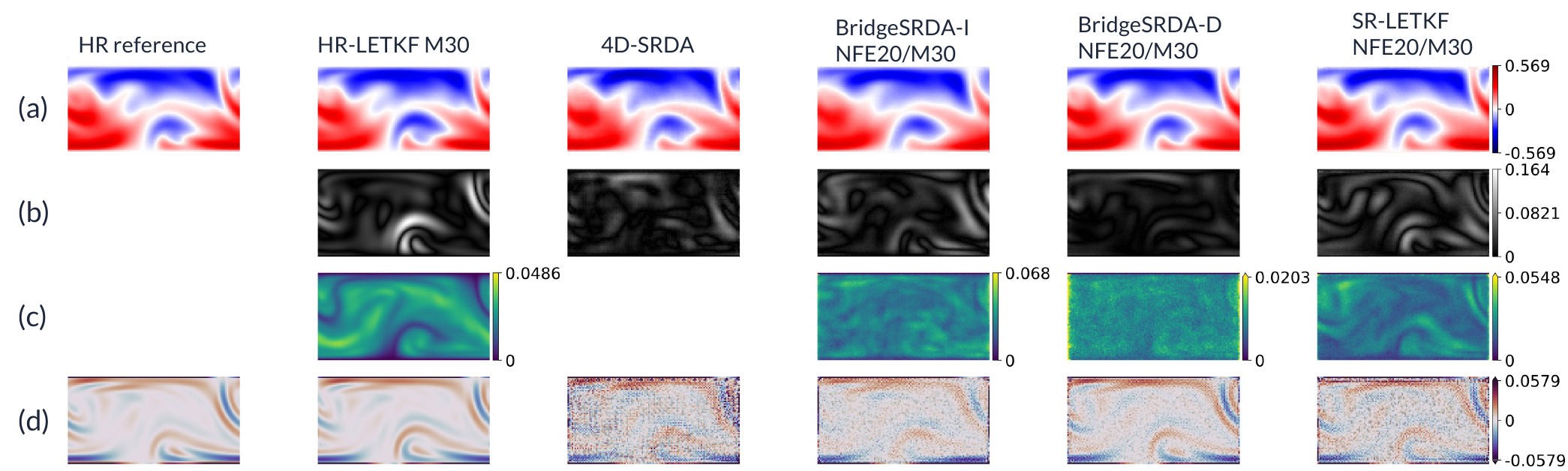} \caption{Analyses at physical time 193 for the bottom layer, zero-based index 0, of the same test trajectory. Columns show the methods and rows show (a) PV, (b) absolute PV error, (c) ensemble spread and (d) unit-grid horizontal Laplacian PV, as in Figure~\ref{fig:analysis-index0500-middle}. A layer-specific Laplacian range makes the smaller bottom-layer curvature visible. Spread uses method-specific ranges, with a colorbar beside each snapshot.}
\label{fig:analysis-index0772-bottom}
\end{figure}

The flow fields in Figures~\ref{fig:analysis-index0500-middle}--\ref{fig:analysis-index0772-bottom} help explain the differences in the aggregate scores. They show analyses at physical times 50, 125 and 193, spanning the development of the instability and the later nonlinear flow.

In the middle layer at time 125 (Figure~\ref{fig:analysis-index0500-middle}a), BridgeSRDA-D reproduces the large-scale PV configuration and much of its finer structure. Localized errors and differences in filament curvature remain (Figure~\ref{fig:analysis-index0500-middle}b,d), consistent with its higher Laplacian-PV RMSE than HR-LETKF. BridgeSRDA-I and SR-LETKF depart more visibly from the reference, in agreement with their larger reconstruction errors.

Figures~\ref{fig:analysis-index0200-top} and \ref{fig:analysis-index0772-bottom} show the top layer at time 50 and the bottom layer at time 193, respectively. They illustrate how the methods reconstruct the flow at different depths and stages of its evolution. The scores in Section~\ref{sec:reconstruction-accuracy} provide the comparison over all analysis times and trajectories.

Although 4D-SRDA captures the broad PV configuration, its Laplacian fields reveal less clearly resolved filament curvature and grid-scale artifacts that are less apparent in the PV fields. The loss of uncertain fine-scale detail is consistent with pointwise regression toward a conditional central estimate. For an $L^1$ objective, this estimate is a conditional median, rather than a set of plausible HR realizations.

The spread maps provide a first view of where ensemble members differ. BridgeSRDA-D shows relatively weak spread overall, together with visible features near the boundaries. The magnitude and reliability of that variability are assessed more directly in the next subsection.

\subsection{Ensemble uncertainty and calibration}\label{sec:results-uq}

\begin{figure}
  \centering
  \includegraphics[width=1.0\textwidth]{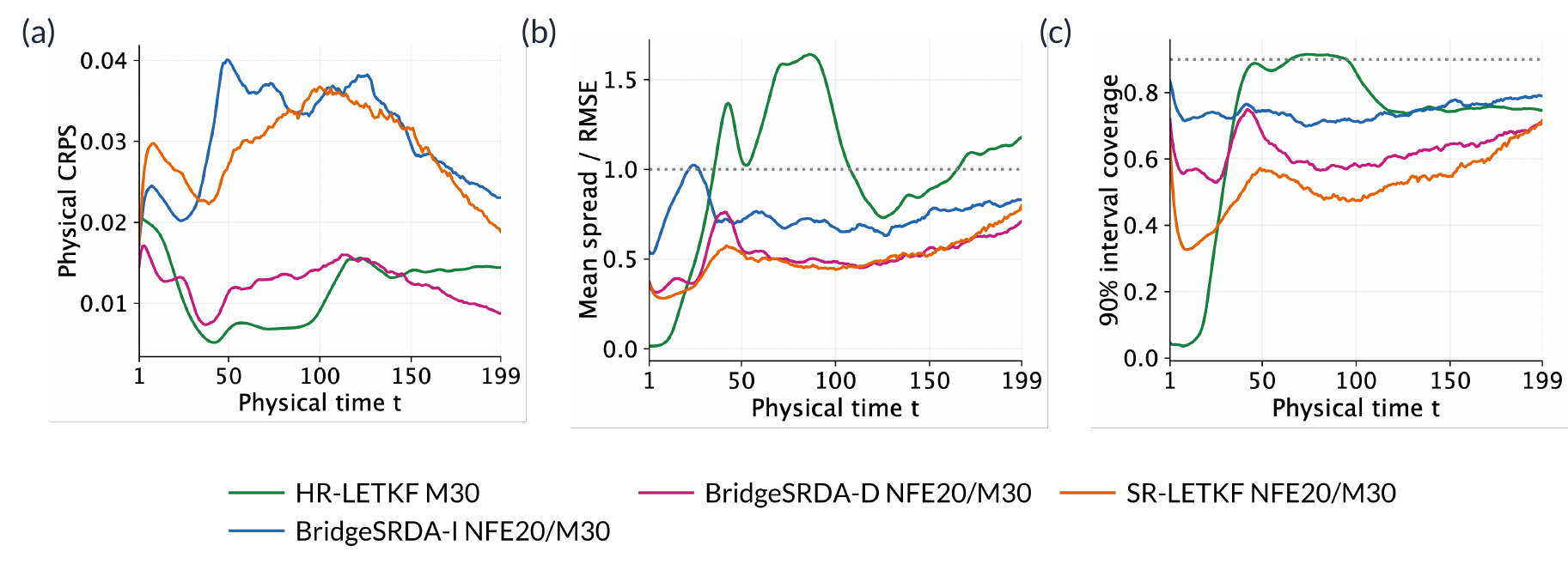} \caption{Ensemble scores averaged over ten trajectories: (a) CRPS in physical PV units, (b) mean per-cycle spread/RMSE and (c) nominal 90\% interval coverage. The reference lines indicate a spread/skill ratio of one and coverage of 0.9. These scores assess the ensemble marginally at individual grid points.}
\label{fig:04-uq}
\end{figure}

BridgeSRDA-D's accurate reconstruction is accompanied by competitive distributional skill (Figure~\ref{fig:04-uq}a and Table~\ref{tab:canonical-ensemble-scores}). Its CRPS is 0.01250, close to 0.01174 for HR-LETKF M30. The other diagnostics, however, show that its ensemble does not fully represent the remaining uncertainty. Its nominal 90\% intervals contain 62.3\% of reference values, and its mean spread/skill ratio is 0.524. Thus, despite its accurate ensemble mean, the spread is too small relative to the reconstruction error.

BridgeSRDA-I shows a different balance. Its ensembles are broader, with a spread/skill ratio of 0.746 and coverage of 74.4\%, indicating less severe underdispersion than BridgeSRDA-D. However, its CRPS is substantially higher at 0.03067, and its ensemble mean is less accurate. Greater coverage in this case therefore does not mean a more accurate reconstruction or a better CRPS.

\begin{table}[!htbp]\centering\small
\caption{Ensemble scores at the observation spacing used during training, averaged over ten trajectories and 199 analyses per trajectory. CRPS is the empirical finite-ensemble score in physical PV units. Spread/skill is the average of per-cycle ratios. All methods use 30 members. Coverage is the fraction of reference values between the ensemble's 5th and 95th percentiles, including boundary rows.}
\label{tab:canonical-ensemble-scores}
\begin{adjustbox}{max width=\linewidth}
\begin{tabular}{lrrr}
\toprule
Method & CRPS & Spread/skill & Coverage90 \\
\midrule
HR-LETKF M30 & 0.01174 & 0.988 & 0.707 \\
BridgeSRDA-D & 0.01250 & 0.524 & 0.623 \\
BridgeSRDA-I & 0.03067 & 0.746 & 0.744 \\
SR-LETKF & 0.02912 & 0.505 & 0.525 \\
\bottomrule
\end{tabular}
\end{adjustbox}
\end{table}

The spread fields also differ in appearance. In Figures~\ref{fig:analysis-index0500-middle}--\ref{fig:analysis-index0772-bottom}, BridgeSRDA-I shows more readily discernible patterns around flow structures, whereas BridgeSRDA-D's spread is weaker and has a more speckled appearance, with features near the boundaries. These visual differences are informative, although they do not by themselves establish whether either ensemble is well calibrated.

BridgeSRDA-D also exhibits locally exaggerated spread near the periodic $x$-boundary, even though it is underdispersed overall. In an additional experiment, a model trained from scratch with circular $x$-padding reduced this boundary concentration but produced less accurate reconstructions. Improving the boundary behavior without losing reconstruction skill remains a topic for future work.

\begin{figure}
  \centering
  \includegraphics[width=0.8\textwidth]{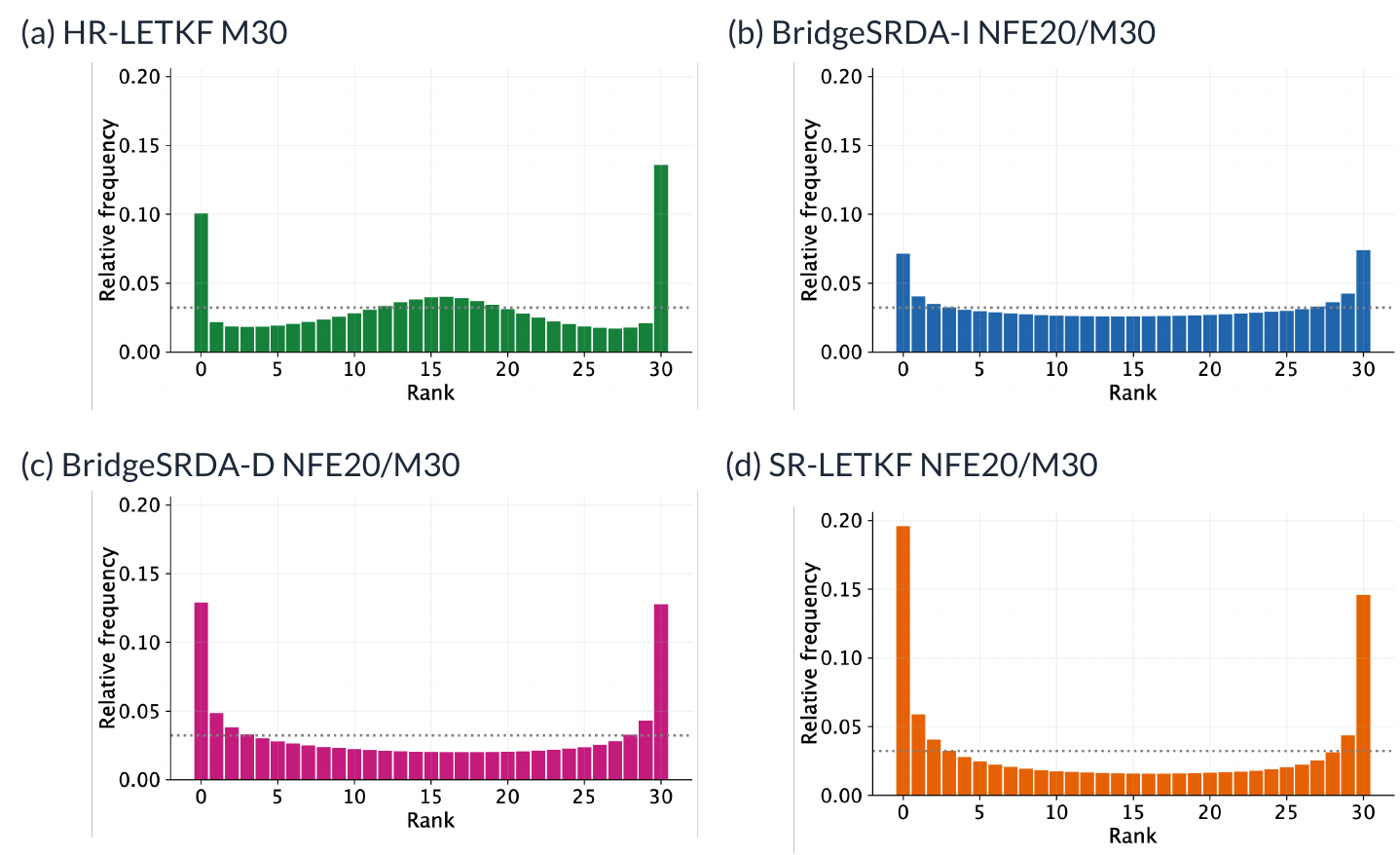} \caption{Pooled rank histograms for (a) HR-LETKF M30, (b) BridgeSRDA-I, (c) BridgeSRDA-D and (d) SR-LETKF. The learned ensembles use NFE20/M30. Concentration at the extreme ranks indicates that the reference often falls outside the ensemble, while a central concentration can indicate excessive dispersion. Bias, spatial heterogeneity and ties can also affect the histogram shape. Ranks use a strict-less-than rule without randomized ties. Boundary conventions are described in Appendix~\ref{app:implementation}.}
\label{fig:04-rank-histograms}
\end{figure}

The rank histograms provide a complementary view (Figure~\ref{fig:04-rank-histograms}). BridgeSRDA-D has a stronger concentration at the extreme ranks than BridgeSRDA-I (Figure~\ref{fig:04-rank-histograms}b,c), consistent with its lower coverage and spread/skill ratio. HR-LETKF also departs from a uniform rank histogram and has 70.7\% coverage, despite an average spread/skill ratio close to one. Thus, an appropriate average spread relative to error does not by itself ensure reliable intervals across the domain and over time.

Together, these results show the importance of assessing reconstruction and uncertainty jointly. BridgeSRDA-D produces accurate analyses and a CRPS close to HR-LETKF, but its intervals are too narrow. BridgeSRDA-I has broader intervals, but substantially larger errors. The diagnostics used here assess individual grid points; they do not establish whether the sampled fields reproduce the full spatial and temporal dependence of the uncertainty.

\subsection{Sampling cost and accuracy}

\begin{figure}
  \centering
  \includegraphics[width=1.0\textwidth]{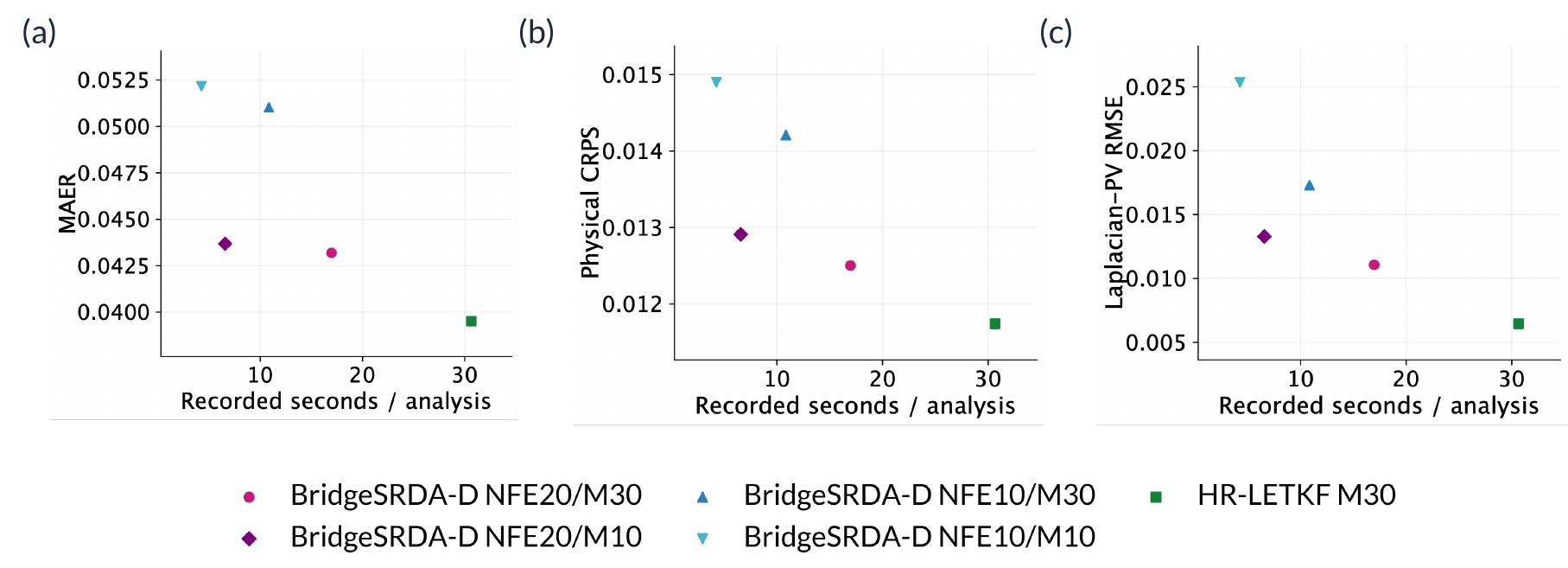} \caption{Recorded cost and (a) MAER, (b) CRPS in physical PV units and (c) horizontal Laplacian-PV RMSE for four BridgeSRDA-D sampling budgets and HR-LETKF M30. Times were measured on an NVIDIA RTX A6000 GPU under potentially different occupancy conditions. Offline training is excluded.}
\label{fig:05-cost-skill}
\end{figure}

We next examine how the sampling budget affects BridgeSRDA-D's accuracy and recorded cost. Figure~\ref{fig:05-cost-skill} and Table~\ref{tab:bridge-budgets} compare four combinations of NFE and ensemble size using the same trained model.

At NFE20, reducing the ensemble from 30 to ten members changes MAER from 0.04319 to 0.04368, an increase of approximately 1.1\%, while reducing the recorded cost from 16.97 to 6.57 s per analysis. These correspond to speed factors of approximately 1.81 and 4.66 relative to the recorded 30.64 s per analysis for HR-LETKF M30. The ten-member ensemble therefore retains strong mean reconstruction at a lower cost, although its Laplacian error increases and its coverage decreases from 0.623 to 0.546.

\begin{table}[!htbp]\centering\small
\caption{Accuracy, uncertainty and recorded cost for four BridgeSRDA-D sampling budgets. All configurations use the same checkpoint and assimilation cadence. NFE counts neural function evaluations per window and member, with two evaluations per Heun interval. M denotes the number of ensemble members.}
\label{tab:bridge-budgets}
\begin{adjustbox}{max width=\linewidth}
\begin{tabular}{lrrrrr}
\toprule
Budget & MAER & Laplacian RMSE & CRPS & Coverage90 & s/analysis \\
\midrule
NFE20/M30 & 0.04319 & 0.01107 & 0.01250 & 0.623 & 16.97 \\
NFE20/M10 & 0.04368 & 0.01328 & 0.01291 & 0.546 & 6.57 \\
NFE10/M30 & 0.05104 & 0.01730 & 0.01421 & 0.765 & 10.85 \\
NFE10/M10 & 0.05216 & 0.02535 & 0.01490 & 0.683 & 4.26 \\
\bottomrule
\end{tabular}
\end{adjustbox}
\end{table}

\begin{figure}
  \centering
  \includegraphics[width=0.6\textwidth]{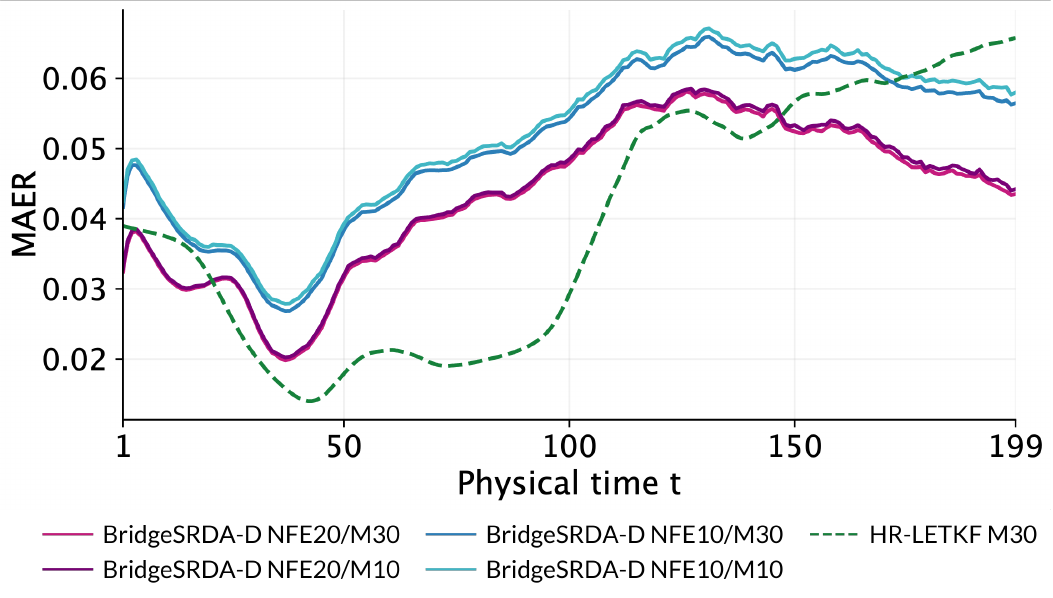} \caption{Ten-trajectory mean MAER during cycling for BridgeSRDA-D with NFE20/M30, NFE20/M10, NFE10/M30 and NFE10/M10. HR-LETKF M30 is included for comparison.}
\label{fig:05-sampling-maer-time}
\end{figure}

Reducing NFE from 20 to ten has a larger effect on reconstruction accuracy than reducing the ensemble size at NFE20. MAER rises to approximately 0.0510--0.0522, and Laplacian error also increases (Table~\ref{tab:bridge-budgets} and Figure~\ref{fig:05-sampling-maer-time}). Coverage, however, increases to 0.765 with 30 members and 0.683 with ten members. The sampling budget therefore affects reconstruction and coverage differently. Because these comparisons use recursive cycling, they do not isolate numerical integration error from changes that accumulate over subsequent analyses.

The deterministic 4D-SRDA baseline is substantially cheaper, with a mean cycling time of 115.94 s per trajectory, or approximately 0.58 s per analysis. It produces one HR window in a single network pass, whereas BridgeSRDA-D uses repeated neural evaluations to generate an ensemble. BridgeSRDA-D improves reconstruction accuracy and supplies ensemble output, but its recorded runtime is higher than that of the deterministic baseline. Its smaller parameter count alone does not make ensemble sampling cheaper.

Further reductions are possible without retraining the model. Appendix~\ref{app:execution-efficiency} examines execution changes on one trajectory while retaining the usual observation-refresh rate. In that experiment, compiled mixed-precision BridgeSRDA-D completes a 30-member forecast--analysis cycle in 5.50 s, compared with 18.10 s for an execution-optimized HR-LETKF M30.

The cost breakdown shows the contribution of the LR forecast: it takes approximately 0.25 s per cycle, compared with 11.09 s for the HR forecast. Neural sampling takes 5.24 s for BridgeSRDA-D, while the LETKF update takes 7.01 s. Smaller BridgeSRDA ensembles further reduce cost while retaining very similar mean PV fields. Their reconstruction and uncertainty tradeoffs are reported separately in the appendix.

Another option is to reduce the observation-refresh rate. Appendix~\ref{app:stride} examines cycling with nowcast feedback, which nearly halves the recorded cycling time with a modest accuracy penalty. This changes the assimilation cadence and is therefore separate from the equal-cadence sampling comparisons above.

\subsection{Short-range nowcasting}\label{sec:nowcast}

As described in Section~\ref{sec:method}, the future frames in each sampled window provide short-range nowcasts in addition to the current analysis. We now evaluate their accuracy and uncertainty over the one-unit forecast horizon. These nowcasts use the LR forecast information already included in the source window, without additional observations or sampling calls. For frame leads $\ell=0,\ldots,4$, the physical horizon is $h=0.25\ell$, ranging from the current analysis to one physical time unit ahead.

\begin{figure}
  \centering
  \includegraphics[width=1.0\textwidth]{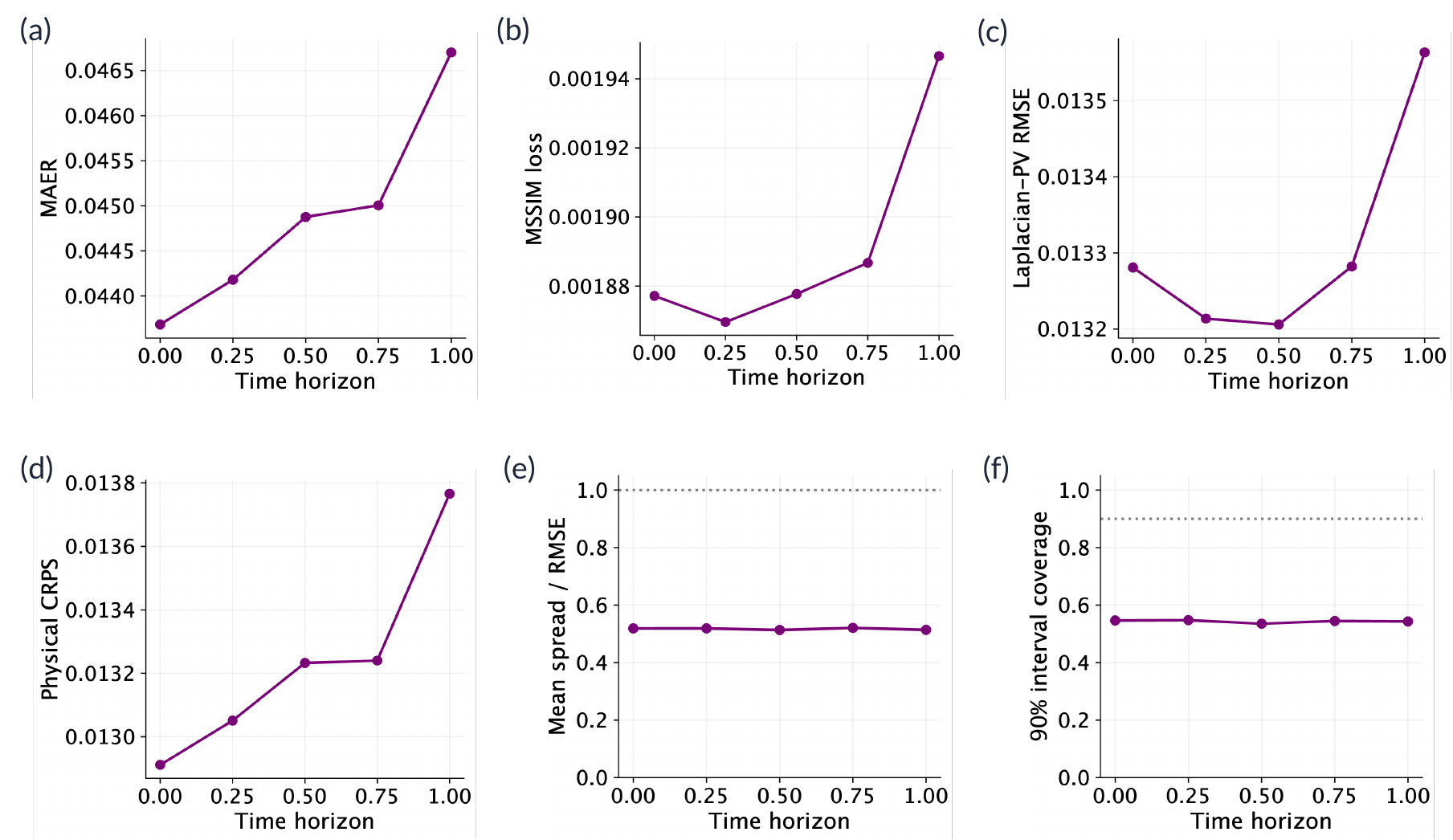} \caption{Reconstruction and uncertainty scores for BridgeSRDA-D NFE20/M10 at physical horizons 0, 0.25, 0.5, 0.75 and 1, averaged over ten trajectories: (a) MAER, (b) MSSIM loss, (c) horizontal Laplacian-PV RMSE, (d) physical CRPS, (e) mean spread/RMSE and (f) 90\% interval coverage. All horizons come from the same sampled window and are measured from the issuance time. Future LR forecasts provide information through the final horizon, but future observations are unavailable.}
\label{fig:06-nowcast}
\end{figure}

Figure~\ref{fig:06-nowcast} and Table~\ref{tab:nowcast-scores} show the results for BridgeSRDA-D with NFE20/M10. MAER is 0.04368 at the current analysis and 0.04670 at the longest horizon, while CRPS is 0.01291 and 0.01377, respectively (Figure~\ref{fig:06-nowcast}a,d). The longest horizon uses one fewer window per trajectory because of the finite reference archive. Coverage remains approximately 0.54--0.55 across the window (Figure~\ref{fig:06-nowcast}f), indicating that the underdispersion seen at the analysis time also persists in the nowcasts.

\begin{table}[!htbp]\centering\small
\caption{Nowcast scores for BridgeSRDA-D with NFE20/M10, averaged over ten trajectories. Leads 0--3 use 199 windows per trajectory, while lead 4 uses 198 because of the finite reference archive. No matched HR-LETKF nowcast experiment is included.}
\label{tab:nowcast-scores}
\begin{adjustbox}{max width=\linewidth}
\begin{tabular}{lrrrr}
\toprule
Physical horizon & MAER & RMSE & CRPS & Coverage90 \\
\midrule
0.00 & 0.04368 & 0.02260 & 0.01291 & 0.546 \\
0.25 & 0.04418 & 0.02281 & 0.01305 & 0.548 \\
0.50 & 0.04488 & 0.02302 & 0.01323 & 0.535 \\
0.75 & 0.04500 & 0.02313 & 0.01324 & 0.545 \\
1.00 & 0.04670 & 0.02403 & 0.01377 & 0.543 \\
\bottomrule
\end{tabular}
\end{adjustbox}
\end{table}

\begin{figure}
  \centering
  \includegraphics[width=1.0\textwidth]{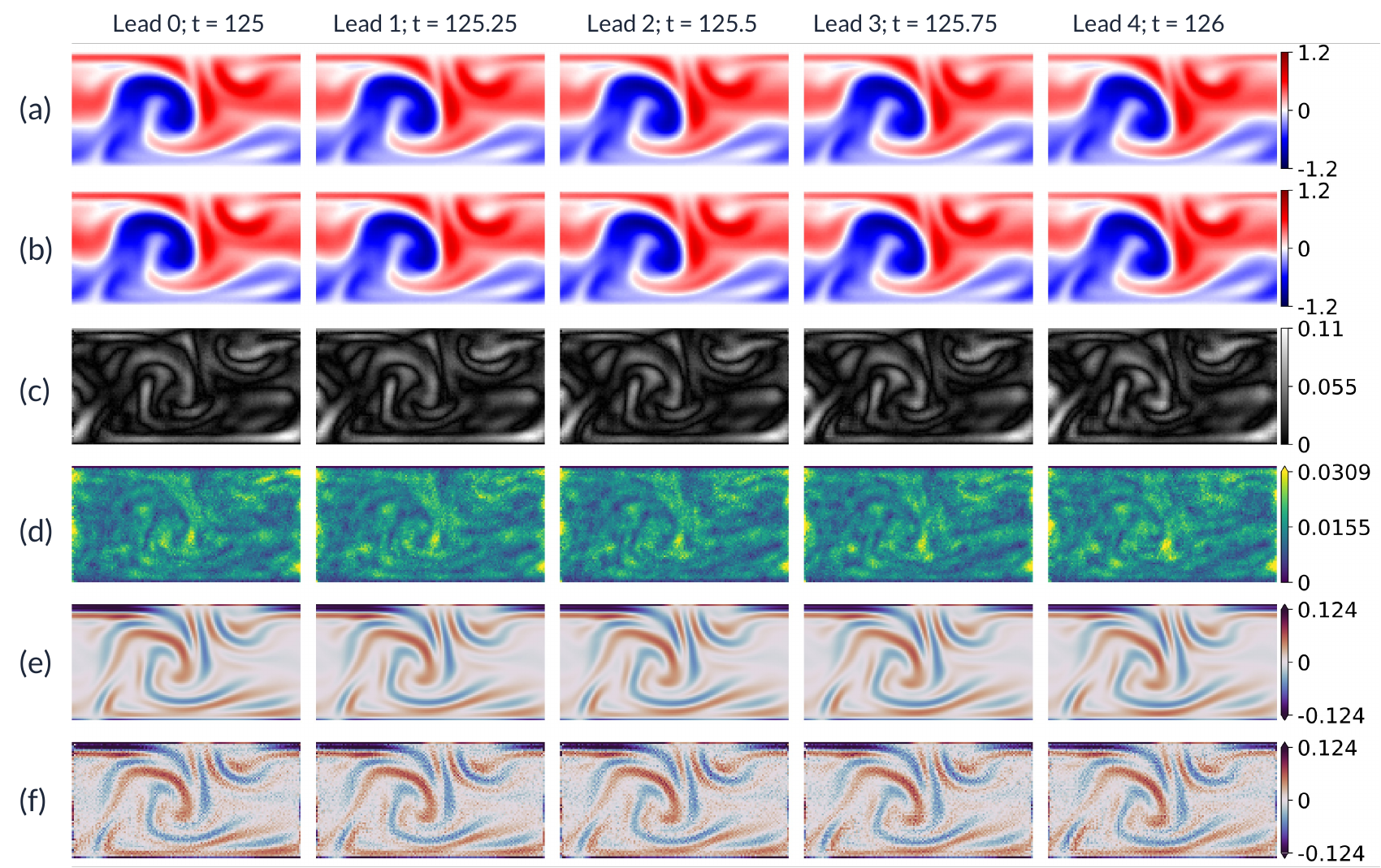} \caption{Nowcasts issued at physical time 125 for the middle layer of one test trajectory. Columns show leads 0--4. Rows show (a) reference PV, (b) predicted mean PV, (c) absolute PV error, (d) ensemble spread, (e) reference Laplacian PV and (f) predicted Laplacian PV. BridgeSRDA-D uses NFE20/M10. Laplacian ranges are matched across leads for this layer. The PV color scale can obscure small temporal changes.}
\label{fig:nowcast-index0500-middle}
\end{figure}

Figure~\ref{fig:nowcast-index0500-middle} shows the issued middle-layer fields from time 125 for one test trajectory. The large-scale flow remains close to the reference over the short horizon. The visible evolution is limited because the physical window is short, and longer windows would provide a more demanding test of prediction as the flow changes.

\subsection{Sensitivity to sparser observations}

We next test whether BridgeSRDA-D can assimilate observations at a spacing not encountered during training. The horizontal lattice spacing is increased from eight to 16 HR grid points in both directions, reducing the number of sensors to approximately one quarter. Observations remain available in every vertical layer. BridgeSRDA-D is applied without retraining, and the LETKF settings are unchanged.

\begin{figure}
  \centering
  \includegraphics[width=1.0\textwidth]{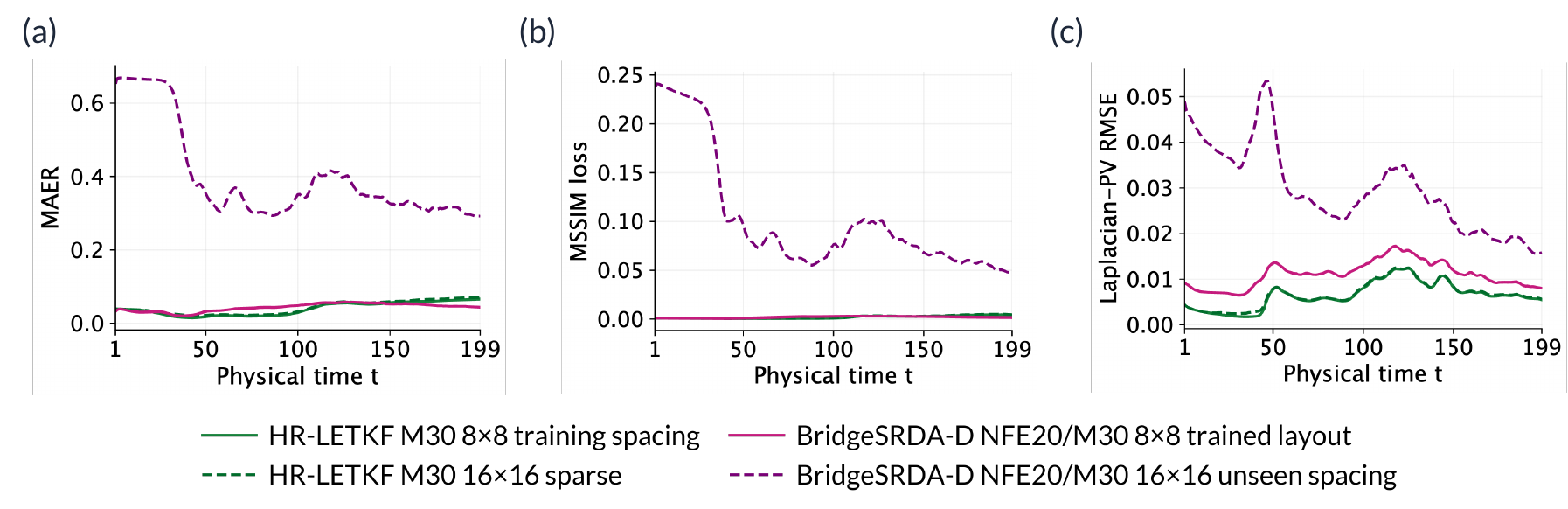} \caption{Reconstruction scores at the training and sparse observation spacings, averaged over ten trajectories, for HR-LETKF M30 and BridgeSRDA-D NFE20/M30: (a) MAER, (b) MSSIM loss and (c) horizontal Laplacian-PV RMSE. Horizontal sensor spacing increases from eight to 16 HR grid points, while all vertical layers remain observed. BridgeSRDA-D was trained at spacing eight with a varying lattice origin. All MSSIM curves use the same CPU implementation.}
\label{fig:10-sparse-letkf-point}
\end{figure}

\begin{figure}
  \centering
  \includegraphics[width=1.0\textwidth]{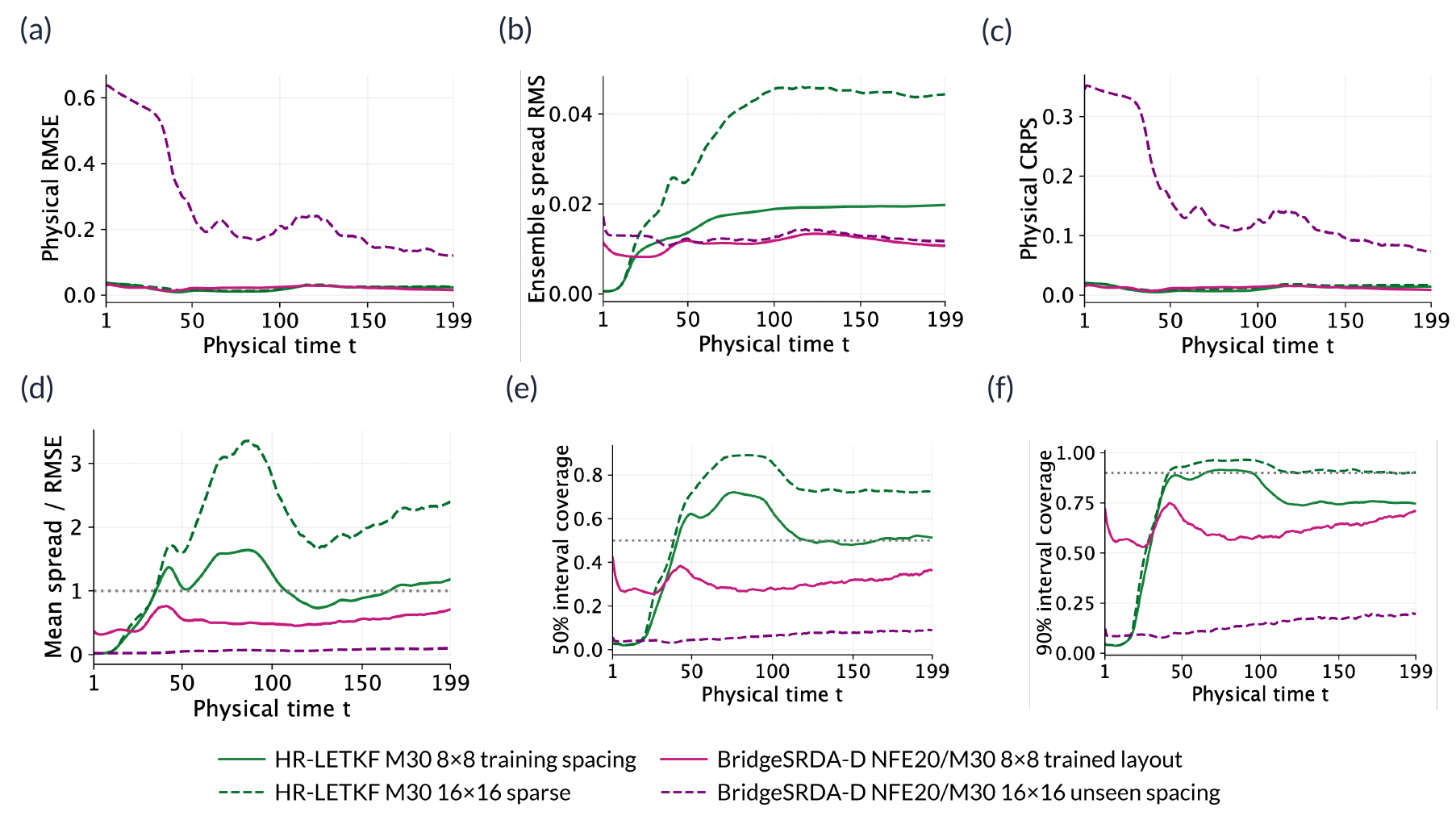} \caption{Ensemble scores at the training and sparse observation spacings, averaged over ten trajectories: (a) physical RMSE, (b) ensemble spread RMS, (c) physical CRPS, (d) mean per-cycle spread/RMSE, and (e) 50\% and (f) 90\% interval coverage. HR-LETKF uses 30 members, and BridgeSRDA-D uses NFE20/M30.}
\label{fig:10-sparse-letkf-uq}
\end{figure}

Figures~\ref{fig:10-sparse-letkf-point} and \ref{fig:10-sparse-letkf-uq} and Table~\ref{tab:sparse-scores} show the resulting scores. HR-LETKF M30 remains accurate at the sparser spacing: its MAER increases from 0.03951 to 0.04292 and its RMSE from 0.02115 to 0.02322. BridgeSRDA-D is much more sensitive to this change. Its MAER rises from 0.04319 to 0.39832, its RMSE rises from 0.02241 to 0.26160, and its 90\% coverage falls from 0.623 to 0.140.

\begin{table}[!htbp]\centering\small
\caption{Scores for HR-LETKF M30 and BridgeSRDA-D NFE20/M30 at the training observation spacing and at the sparser spacing. Values are averaged over ten trajectories, with the model and filter settings unchanged. MSSIM losses are recomputed using the same central processing unit (CPU) implementation for all four cases and differ slightly from those in Table~\ref{tab:canonical-point-scores}. The two spacing scenarios use independently generated observation-noise realizations. The LETKF observation-covariance convention remains as described in Section~\ref{sec:comparison-methods}.}
\label{tab:sparse-scores}
\begin{adjustbox}{max width=\linewidth}
\begin{tabular}{lrrrrr}
\toprule
Method/scenario & MAER & MSSIM loss & RMSE & CRPS & Coverage90 \\
\midrule
HR-LETKF: trained spacing & 0.03951 & 0.001764 & 0.02115 & 0.01174 & 0.707 \\
HR-LETKF: sparse & 0.04292 & 0.002020 & 0.02322 & 0.01429 & 0.805 \\
BridgeSRDA-D: trained spacing & 0.04319 & 0.001867 & 0.02241 & 0.01250 & 0.623 \\
BridgeSRDA-D: sparse & 0.39832 & 0.101756 & 0.26160 & 0.15722 & 0.140 \\
\bottomrule
\end{tabular}
\end{adjustbox}
\end{table}

\begin{figure}
  \centering
  \includegraphics[width=1.0\textwidth]{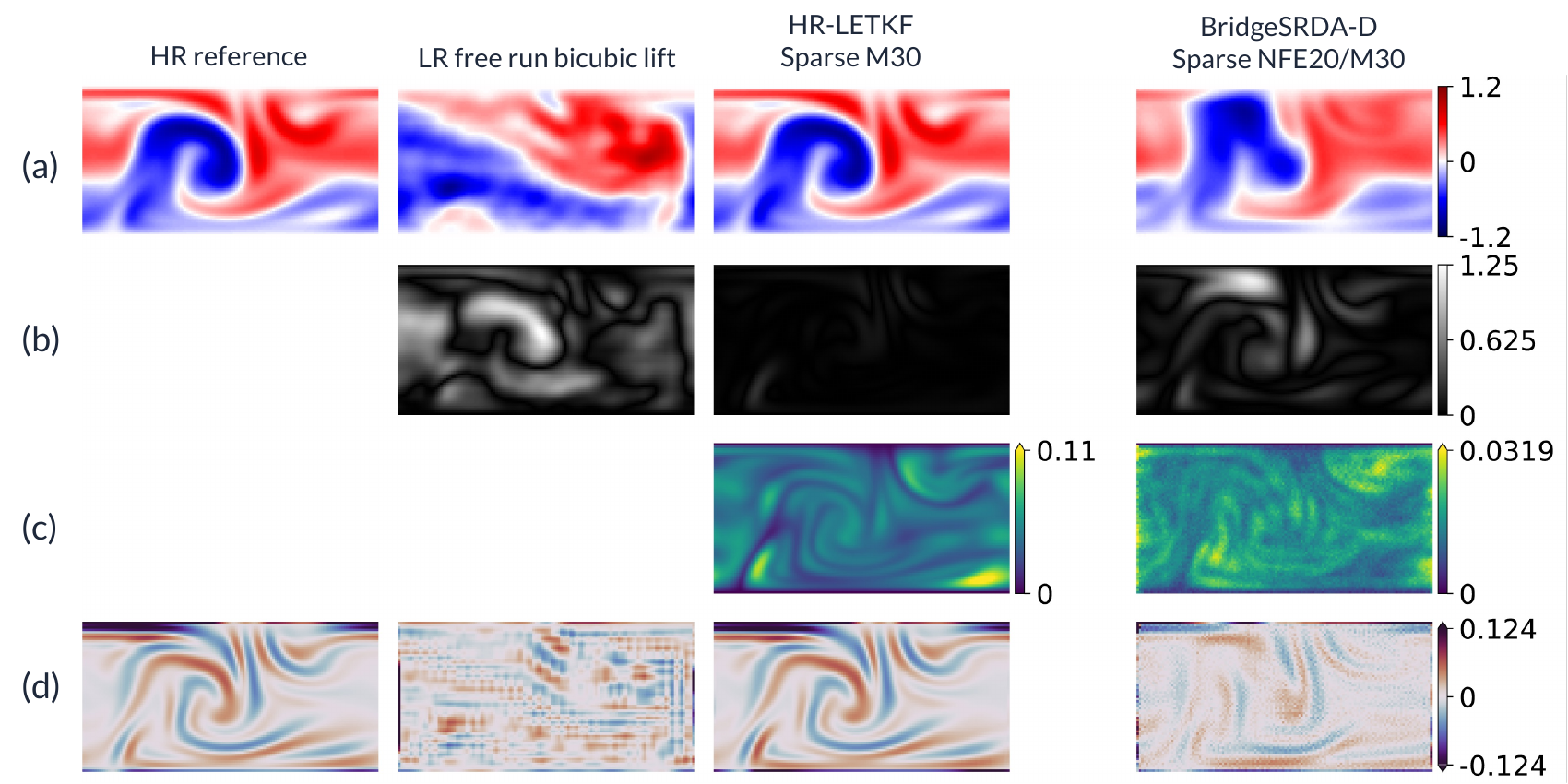} \caption{Sparse-observation analyses from HR-LETKF M30 and BridgeSRDA-D NFE20/M30, alongside HR truth and an unassimilated bicubic-lifted LR free run. The fields are shown at physical time 125 for the middle layer of one test trajectory. Rows show (a) PV, (b) absolute PV error, (c) ensemble spread and (d) Laplacian PV. Spread uses method-specific ranges, with a colorbar beside each snapshot, so equal colors do not imply equal spread magnitudes.}
\label{fig:sparse-index0500-middle}
\end{figure}

The deterioration in BridgeSRDA-D is not accompanied by a corresponding increase in ensemble spread (Figure~\ref{fig:10-sparse-letkf-uq}a,b). Mean spread rises only from 0.01126 to 0.01259, while RMSE increases by more than an order of magnitude. The ensemble therefore does not adequately signal the loss of accuracy. Figure~\ref{fig:sparse-index0500-middle} shows that, for one test case, the sparse-observation reconstruction retains some large-scale agreement with the reference that is absent from the LR free run, but its fine structures are poorly recovered.

Several features of this benchmark help provide information to HR-LETKF even at the sparser spacing. PV is observed in every vertical layer, updates are frequent, and the forecast and nature models share their governing dynamics despite their different resolutions. LETKF incorporates the available measurements through its observation operator and covariance update. BridgeSRDA-D, by contrast, receives interpolated conditions whose distribution changes when the sensor spacing differs from training.

The sparse-observation experiment identifies an important limitation of the current training design. BridgeSRDA-D performs well at the spacing for which it was trained, but it does not yet adapt reliably to the sparser layout. Training across observation layouts and developing representations that account for sensor support are natural directions for improving both the analysis and its uncertainty.

\section{Further discussion}\label{sec:discussion}

\subsection{Conditional diffusion experiments in three dimensions}

In our previous DiffSRDA study, we used conditional DDPM-based generation for probabilistic SRDA with a two-dimensional barotropic jet dataset~\citep{Ayapilla2026DiffSRDA}. That result made conditional diffusion a natural starting point for the present three-dimensional problem. We trained DDPM-style models to predict injected Gaussian noise and used respaced sampling from the denoising diffusion implicit model (DDIM) family~\citep{Song2021DDIM}.

The tested diffusion models did not produce useful HR reconstructions under the training and sampling settings we explored. After 125 epochs, the reconstruction from an approximately 89.54-million-parameter diffusion U-Net remained strongly distorted, whereas BridgeSRDA-D had already learned recognizable HR structure with a similar parameter count. For a separate MaxViT diffusion model, increasing the sampling budget to 100--500 evaluations did not recover useful reconstruction. Appendix~\ref{app:diffusion} presents these experiments and their validation fields.

These results motivated us to investigate a different generative formulation. The comparison does not isolate dimensionality or the choice of generative model alone, since conditioning, training exposure and validation inputs also differ. It therefore does not rule out a successful diffusion-based approach for the present problem. It does show that the straightforward extensions we tested were insufficient within the explored budgets, while forecast-informed stochastic transport provided a workable alternative for sequential SRDA.

\subsection{Why forecast-informed transport may help SRDA}

The LR forecast provides an organized approximation to the flow before the neural reconstruction begins. BridgeSRDA uses that forecast both as the initial state of the transport and as a condition throughout sampling. The network can therefore use the available flow structure while learning how to account for unresolved scales, forecast errors and observations.

This starting point offers a possible advantage over generating the entire HR field from unstructured noise. In a two-dimensional turbulence-SR study, \citet{Schiodt2026TurbulenceSI} found that stochastic-interpolant models initialized from low-resolution fields reproduced the distribution of spatially averaged dissipation with fewer sampling steps than the diffusion and flow-matching models considered. Our experiments extend the investigation of such an informative source to observation-conditioned reconstruction within a sequential DA cycle.

The LR source carries information produced by the governing dynamics, but it cannot resolve all fine-scale interactions. The HR analysis is generated using the learned model and observations, so it is not guaranteed to satisfy the governing equations exactly. Its usefulness in cycling depends on whether it provides an accurate enough state to initialize the next numerical forecast.

The drift target provides another way to understand the reconstruction task. If we write the paired HR target as $Y_1=Y_0+R$, the correction $R$ contains the differences between the lifted forecast and its HR target, including phase errors, forecast-model errors and unresolved scales. The supervised drift can then be separated into a part determined by the source and transport time, a part associated with this correction, and a stochastic contribution. Appendix~\ref{app:brownian} gives the decomposition. The network still predicts the full drift. It is not trained as an explicitly residualized model.

Under squared-error training, an ideal network estimates the mean target drift conditional on the intermediate state, forecast source and observations. For the p2 construction, this drift can also be expressed in terms of the conditional mean HR endpoint, as shown in Equation~\eqref{eq:drift-endpoint-mean}. This relationship helps explain how drift learning connects an intermediate state to HR outcomes compatible with the supplied conditions. It does not by itself guarantee successful learning from finite data or calibrated ensembles during cycling.

The noise schedule and sampler complement the forecast-informed source. The quadratic p2 envelope reduces the stochastic forcing near the HR endpoint and makes the explicit random contribution to the interior drift target vanish there. It retains the same total sampling-noise variance as the p1 envelope used for comparison. These properties motivate a cleaner final stage of reconstruction.

The p1/p2 validation results in Appendix~\ref{app:noise} support the selected p2 recipe, although they do not isolate the effect of the envelope from every difference in training execution. Likewise, continuous stratified time sampling and Heun integration are parts of the chosen method, but their separate contributions were not measured in isolation. The results therefore support the complete reconstruction framework rather than establishing which individual choice is responsible for its performance.

\subsection{The distribution learned by BridgeSRDA}

As described in Section~\ref{sec:method}, each BridgeSRDA cycle produces a current analysis, reconstructions of earlier states and short-range nowcasts. Observations available at issuance can inform the earlier frames, giving the window a smoothing role, while the current analysis is used to continue the forecast--analysis cycle. These roles describe how BridgeSRDA uses the available information. They do not imply that it performs exact Bayesian filtering or smoothing.

The distribution learned during training is determined by the paired HR targets and the forecast and observation conditions supplied to the network. Under ideal drift regression, matching diffusion, continuous integration and suitable regularity conditions, the construction reproduces the corresponding conditional distribution of HR targets. In deployment, finite training and sampling errors, together with changes in the forecast sources, can affect that correspondence.

Even if the training distribution were learned exactly, it would not automatically be the filtering posterior conditioned on the complete observation history. As \citet{Hodyss2025DiffusionDA} discuss, generative DA systems can target different posterior distributions depending on their priors and conditioning. Accurate reconstruction alone is therefore not enough to establish that a learned system reproduces the posterior of a fully cycled Bayesian filter.

BridgeSRDA carries information from previous observations in two ways. Past and current measurements condition the network explicitly, and previous generated analyses initialize the member-wise LR forecasts. This allows information from earlier cycles to influence the current reconstruction. In a linear setting, \citet{Morzfeld2026ObservationHistory} show that conditioning on a sufficiently long observation window can recover the performance of a cycling filter without retraining at every cycle. Whether the information available to BridgeSRDA is sufficient for the nonlinear QG system remains an open question.

There is also a difference between the forecast sources seen during training and cycling. Offline training uses LR forecasts initialized from filtered HR truth, whereas deployment uses forecasts initialized from generated analyses. The resulting change in forecast errors is one possible contributor to the observed underdispersion. Other factors, including finite training data, neural approximation, numerical sampling and the information available to the network, may also matter. Distinguishing these effects will help identify how to improve uncertainty without losing reconstruction accuracy.

\subsection{Computational flexibility in cycling assimilation}

The efficiency experiment in Appendix~\ref{app:execution-efficiency} helps explain where BridgeSRDA's computational savings arise. In that experiment, the LR forecast is much cheaper than the HR forecast, while batching, compilation and mixed precision reduce the cost of neural sampling. At the same 30-member budget, the complete BridgeSRDA-D cycle is approximately 3.3 times faster than the optimized HR-LETKF cycle on the tested GPU. This result illustrates the value of using an inexpensive numerical forecast to support HR analysis without evolving every member numerically on the HR grid.

Ensemble size plays different roles in the two methods. LETKF estimates forecast covariances from its ensemble, so changing the number of members also changes the information used in the analysis update. BridgeSRDA learns its conditional drift during training and samples from it during deployment. Reducing the number of samples changes the resolution of the resulting ensemble. Because each generated member also initializes a subsequent forecast, the change can influence later cycles as well.

In the main comparison, reducing HR-LETKF from 30 to ten members increases MAER by approximately 20\%, whereas the corresponding increase for BridgeSRDA-D is approximately 1.1\%. The separate efficiency experiment also finds small changes in mean reconstruction with five BridgeSRDA members. This provides a useful option when a rapid estimate of the mean flow is the priority.

A small change in mean error does not, however, imply that uncertainty is equally well represented. With fewer members, empirical intervals and their tails are estimated from fewer samples, and the five-member experiment also shows increased Laplacian error. Finite-ensemble effects influence the measured coverage, so changes in coverage alone do not establish how the underlying learned distribution has changed. Applications that require reliable probabilities must therefore assess the ensemble size alongside calibration and fine-scale accuracy.

The reported costs depend on the tested hardware and implementations. Within that setting, the results demonstrate several distinct ways to adjust deployment cost without retraining: improving execution, changing the number of sampled members, or reducing the observation-refresh rate. Each choice has different consequences for accuracy, uncertainty and the information available at each cycle. Their value will depend on the requirements of the application.

\section{Concluding remarks}\label{sec:conclusions}

\subsection{Conclusions}

High-resolution ensemble data assimilation requires repeated forecasts and updates of many possible flow states. As the state dimension increases, the cost of these cycles can become substantial. In this study, we investigated whether a generative model can reduce that cost while retaining accurate high-resolution analyses and useful estimates of uncertainty.

Our approach, BridgeSRDA, combines low-resolution numerical forecasting with observation-conditioned stochastic transport. A neural drift trained using a conditional stochastic interpolant transports the lifted forecasts, conditioned on sparse observations, to generate an ensemble of high-resolution windows. The current analysis from each member initializes its next low-resolution forecast. Each window also provides reconstructions of earlier states and short-range nowcasts.

We evaluated the framework on ten narrow-jet trajectories of a multilayer quasi-geostrophic system. At the observation spacing used during training, BridgeSRDA-D approached HR-LETKF M30 in mean and structural reconstruction accuracy. It reduced MAER by approximately 32.5\% relative to deterministic 4D-SRDA while using approximately 76.3\% fewer network parameters and generating an ensemble of analyses. BridgeSRDA-D also recovered finer-scale structure more accurately than the deterministic model, although HR-LETKF retained an advantage in reconstructing the smallest-scale curvature.

The recorded cost of BridgeSRDA-D with 30 members was approximately 17 s per analysis, compared with 31 s for HR-LETKF M30 under the measured execution conditions. Reducing the BridgeSRDA ensemble to ten members substantially lowered the cost with only a small change in mean reconstruction error, although fine-scale accuracy and interval coverage changed more noticeably.

The efficiency experiment showed that the recorded times are not fixed limits. After execution improvements to both methods, BridgeSRDA-D was approximately 3.3 times faster than HR-LETKF at the same 30-member budget on one tested trajectory. A major contribution to this difference was the numerical forecast: the LR forecast took approximately 0.25 s per cycle, compared with 11.09 s for the HR forecast. Smaller BridgeSRDA ensembles reduced the cost further. These results demonstrate flexibility in choosing a computational budget, while also showing that mean accuracy, fine-scale reconstruction and uncertainty must be considered together.

The uncertainty evaluation revealed both encouraging results and important limitations. BridgeSRDA-D achieved a CRPS close to that of HR-LETKF M30, but its ensemble spread was too small relative to its errors, and localized spread artifacts appeared near the boundaries. Its reconstruction and coverage also deteriorated substantially when the observation spacing was doubled without retraining. These findings show that accurate analyses do not automatically imply reliable uncertainty or robustness to changes in the observing network.

Overall, the results show that inexpensive low-resolution forecasts can support accurate high-resolution ensemble assimilation when combined with a suitably trained generative reconstruction model. This study provides an initial assessment of forecast-informed stochastic transport for sequential SRDA and identifies both its computational flexibility and the challenges that remain in uncertainty calibration and observation-layout robustness. These findings provide a foundation for extending the approach toward more demanding forecasting applications.

\subsection{Future work}\label{sec:future}

Improving the reliability of the sampled uncertainty is an important next step. Further work should distinguish errors associated with the learned distribution, numerical sampling and recurrent cycling, while also addressing the boundary artifacts observed in the spread fields. Such improvements should be assessed alongside reconstruction accuracy and cost rather than through any single metric.

The sensitivity to sparser observations also calls for methods that can adapt to changes in sensor location, density and availability. Training across a wider range of observing layouts and developing observation representations that account for sensor support are possible directions. Posterior guidance offers a complementary approach. For example, \citet{Mucke2026ObservationInterpolants} develop an observation-interpolant framework for conditioning generative models on measurements. How such ideas can improve robustness and calibration in forecast-informed SRDA remains to be investigated.

Finally, longer reconstruction windows and more demanding observation and model settings would test the framework beyond the present QG benchmark. Applications such as localized heavy-rainfall forecasting and urban micrometeorology provide a longer-term motivation, particularly where existing numerical solvers and multiscale forecasts remain valuable. The achievable accuracy, uncertainty and computational savings in such systems will depend on their observations, dynamics and implementation, and must be established through application-specific evaluation.

\section*{Data availability}

This work's source code is available at \url{https://github.com/Pranithhazard/BridgeSRDA}. The datasets and trained model weights required to reproduce the experiments are listed in the repository, but are not currently included in it. They can be made available upon reasonable request.

\section*{Acknowledgements}

This work was supported by the JSPS KAKENHI (Grant Number 25H00715) and Central Japan Railway Company. Some of the simulations and deep learning experiments were performed on the TSUBAME4.0 supercomputer provided by Institute of Science Tokyo and the Earth Simulator at the Japan Agency for Marine-Earth Science and Technology (JAMSTEC).

\FloatBarrier
\bibliographystyle{plainnat}
\bibliography{references}

\clearpage\appendix
\section{Brownian interpretation and bridge optimality}\label{app:brownian}

This appendix explains how the drift target used to train BridgeSRDA follows from a Brownian interpretation of the stochastic interpolant. It also relates the drift to the correction between the LR forecast and its HR target.

As introduced in Section~\ref{sec:method}, a training state at transport time $\tau$ is generated using the Gaussian term $\sqrt{\tau}\,g(\tau)Z$, where $Z\sim\mathcal N(0,\mathbf I)$. For any fixed $\tau$, the random variable $\sqrt{\tau}Z$ has the same distribution as a Wiener process $W_\tau$ evaluated at that time. We can therefore consider the Brownian process
\begin{equation}
\widehat I_\tau=(1-\tau)Y_0+\tau^2Y_1+g(\tau)W_\tau,
\label{eq:brownian-state}
\end{equation}
where $W_\tau$ is independent of the paired windows and observation conditions.

Applying It\^o's product rule gives
\begin{equation}
d\widehat I_\tau=(-Y_0+2\tau Y_1+g'(\tau)W_\tau)d\tau+g(\tau)dW_\tau.
\label{eq:brownian-drift}
\end{equation}
The coefficient of $d\tau$ is the Brownian drift. At a sampled transport time, replacing $W_\tau$ in this expression by a random variable distributed as $\sqrt{\tau}Z$ gives the drift target $B_\tau$ used in Section~\ref{sec:method}.

The Gaussian and Brownian expressions have the same distribution at each fixed time, but a single fixed $Z$ does not trace out a Brownian path as $\tau$ varies. Differentiating the Gaussian expression while holding $Z$ fixed therefore gives an ordinary derivative that differs from the It\^o drift above.

The network is trained to predict $B_\tau$ from the intermediate state, source, observations and transport time. Under squared-error regression, the ideal predictor is
\begin{equation}
v^*(x,Y_0,C_{\rm obs},\tau)=\mathbb E[B_\tau\mid I_\tau=x,Y_0,C_{\rm obs},\tau].
\label{eq:conditional-drift}
\end{equation}
In other words, the ideal network predicts the average drift among training examples compatible with its inputs. Different HR targets and noise realizations can contribute to that average.

With this ideal drift and the matching diffusion coefficient $g(\tau)$, the continuous sampling process has the conditional marginals prescribed by the interpolant, under suitable regularity conditions~\citep{Chen2024ConditionalInterpolants}. In practice, the learned network, numerical integration and differences between training and deployment inputs can affect this correspondence.

Matching the prescribed conditional marginals is distinct from establishing the optimality of a Schr\"odinger bridge relative to a specified reference process. \citet{Chen2024ConditionalInterpolants} also describe a special F\"ollmer construction with an adjusted drift and diffusion. For the coefficients used here, its squared diffusion coefficient is
\begin{equation}
\left(g_\tau^{\rm F}\right)^2=\epsilon^2(1-\tau)^3(3+\tau).
\end{equation}
BridgeSRDA instead uses the matching diffusion
\begin{equation}
g(\tau)^2=\epsilon^2(1-\tau)^4.
\end{equation}
We therefore do not assign the specific F\"ollmer optimality result to the implemented BridgeSRDA sampler.

\subsection{Forecast correction and the conditional endpoint mean}

Each training pair consists of a lifted LR forecast window $Y_0$ and its HR target $Y_1$. We can write the target as
\begin{equation}
Y_1=Y_0+R,\qquad R=Y_1-Y_0.
\end{equation}
Here, $R$ is the correction needed to turn that particular forecast into its paired target. It can include differences in phase and amplitude as well as unresolved fine-scale structure.

Substituting $Y_1=Y_0+R$ into Equations~\eqref{eq:interpolant} and \eqref{eq:drift-target} gives, for the quadratic envelope $p=2$,
\begin{equation}
\begin{aligned}
I_\tau&=(1-\tau+\tau^2)Y_0+\tau^2R
          +\epsilon(1-\tau)^2\sqrt{\tau}Z,\\
B_\tau&=(2\tau-1)Y_0+2\tau R
          -2\epsilon(1-\tau)\sqrt{\tau}Z.
\end{aligned}
\label{eq:forecast-correction}
\end{equation}
This expression separates the target drift into three parts. The first depends on the source $Y_0$ and transport time. The second depends on the forecast-to-target correction $R$. The third is the stochastic contribution.

This decomposition is an interpretation of the existing training target. The HR correction $R$ is not available during deployment, and the network is not explicitly trained to predict only that correction. It predicts the full drift without subtracting the source-dependent term during training or adding it back during sampling.

At $\tau=1/2$, the source-only term vanishes, giving
\begin{equation}
B_{1/2}=R-\frac{\epsilon Z}{\sqrt{2}}.
\end{equation}
At this transport time, the drift target consists of the paired forecast correction $R$ together with a stochastic contribution. The intermediate training state itself is formed from weighted contributions of the forecast, the HR target and noise. For example, at $\tau=1/2$, the forecast has weight $1/2$ and the target has weight $1/4$.

The network learns the transformation using intermediate windows sampled at different values of $\tau$. Here, $\tau$ measures progress through the generative transformation, while the frames within each window correspond to physical times. The interpolant does not require these intermediate windows to satisfy the governing fluid equations.

The ideal drift can also be related to the HR endpoints compatible with a given intermediate state. For $0<\tau<1$, define their conditional mean as
\begin{equation}
m_\tau(x,Y_0,C_{\rm obs})
  =\mathbb E[Y_1\mid I_\tau=x,Y_0,C_{\rm obs},\tau].
\end{equation}
Eliminating $Z$ between the interpolant and drift-target equations then gives
\begin{equation}
v^*(x,Y_0,C_{\rm obs},\tau)
  =Y_0+\frac{2}{1-\tau}\bigl[\tau m_\tau(x,Y_0,C_{\rm obs})-x\bigr].
\label{eq:drift-endpoint-mean}
\end{equation}
This identity relates the ideal drift to the mean of the HR targets compatible with the intermediate state and its conditions.

The expression above is derived for interior transport times and has a singular denominator at $\tau=1$. During training, the endpoint targets instead follow directly from the original drift-target equation:
\begin{equation}
B_0=-Y_0,\qquad B_1=-Y_0+2Y_1.
\end{equation}
During sampling, the network predicts the drift. The true HR target $Y_1$ is not available to the sampler.

\section{Noise-envelope comparison}\label{app:noise}

\subsection{Effects of the noise envelope}

Section~\ref{sec:method} introduced the quadratic noise envelope used by BridgeSRDA. Here we compare it with a linear envelope to explain how the two schedules distribute noise over transport time.

Both envelopes have the form
\begin{equation}
g_p(\tau)=\epsilon_p(1-\tau)^p.
\end{equation}
We use $p=1$ for the linear envelope, denoted p1, and $p=2$ for the quadratic envelope, denoted p2. Their amplitudes are
\begin{equation}
\epsilon_1=0.2,\qquad \epsilon_2=0.2\sqrt{\frac{5}{3}}.
\end{equation}
These choices give both schedules the same total sampling-noise variance over the transport interval:
\begin{equation}
\int_0^1 g_p(\tau)^2\,d\tau
=\frac{\epsilon_p^2}{2p+1}\approx0.0133333.
\end{equation}
The schedules therefore introduce the same total sampling-noise variance, but distribute it differently over transport time.

The envelope also affects the explicit noise contribution to the drift target used during training. For a fixed pair of endpoints, the variance of this contribution in one coordinate is
\begin{equation}
\operatorname{Var}_Z(B_{\tau,i}\mid Y_0,Y_1,\tau)
   =p^2\epsilon_p^2\tau(1-\tau)^{2p-2},\qquad 0<\tau<1.
\label{eq:drift-noise-variance}
\end{equation}
As $\tau$ approaches one from within the interval, this variance approaches $\epsilon_1^2$ for p1 and zero for p2. At the exact endpoint, both implementations use the drift target with the explicit noise contribution marginalized out.

For the matched amplitudes above, the ratio of p2 to p1 drift-target noise variance is
\begin{equation}
\frac{20}{3}(1-\tau)^2.
\end{equation}
At $\tau=0.9$, the p2 variance is one fifteenth of the p1 value. Earlier in the interval, their relationship differs, and p2 does not have lower variance at every transport time. This calculation describes the explicit random term for fixed endpoints. The uncertainty in the learned drift and the spread of generated ensemble members depend on additional factors.

During sampling, the noise envelope determines how much new stochastic variance is introduced as the state moves toward the endpoint. The variance remaining after time $\tau$ is
\begin{equation}
V_p(\tau,1)=\int_\tau^1 g_p(s)^2\,ds
   =\frac{\epsilon_p^2}{2p+1}(1-\tau)^{2p+1}.
\label{eq:remaining-noise-variance}
\end{equation}
For the matched schedules and $\tau<1$,
\begin{equation}
\frac{V_2(\tau,1)}{V_1(\tau,1)}=(1-\tau)^2.
\end{equation}
Thus, during the final tenth of transport, p2 introduces one hundredth of the sampling-noise variance introduced by p1. Both variances reach zero at the endpoint.

\begin{figure}
  \centering
  \includegraphics[width=1.0\textwidth]{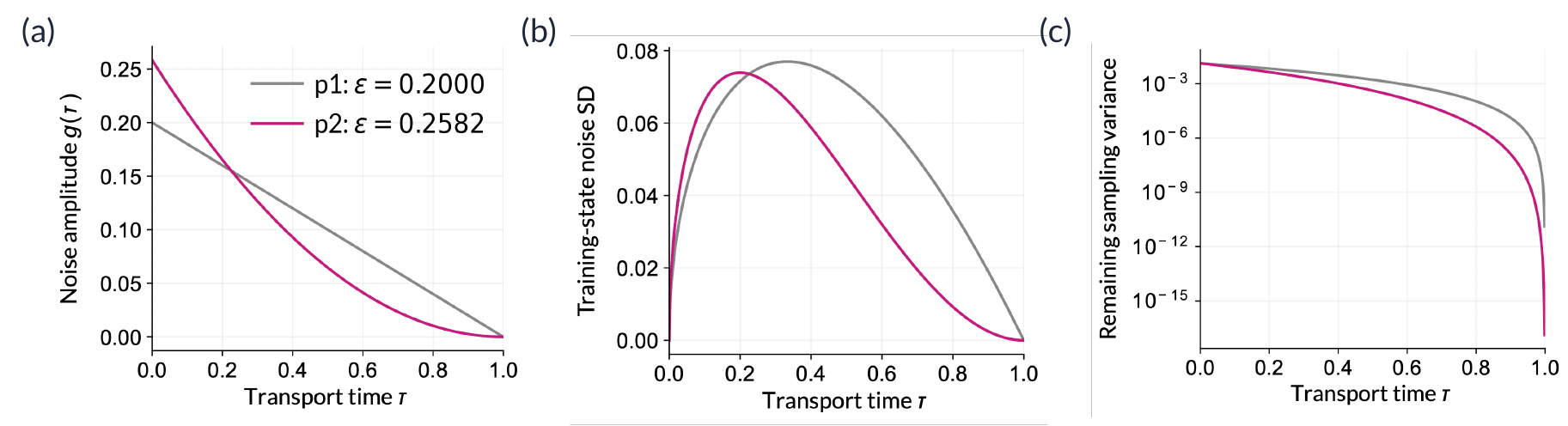} \caption{Linear p1 and quadratic p2 noise envelopes with equal total sampling-noise variance. Panels show (a) noise amplitude, (b) standard deviation of the explicit noise contribution to the training state, and (c) remaining sampling-noise variance. For fixed endpoints, the training-state noise variance is $\tau g(\tau)^2$. The remaining sampling variance describes noise added after $\tau$. Panel (c) uses a logarithmic scale and omits the endpoint, where the variance is zero.}
\label{fig:noise-schedule-2}
\end{figure}

Figure~\ref{fig:noise-schedule-2} shows the two envelopes, the standard deviation of their explicit training-state noise and the sampling variance remaining at each transport time. The quadratic envelope places more of the sampling noise earlier in the transport, leaving a quieter final stage as the state approaches the HR endpoint.

\subsection{Observation-free SR validation}

We next compare the two schedules using the saved validation results for observation-free SR models. These models reconstruct HR windows from LR forecasts without observation conditions. The comparison therefore examines the SR training and sampling recipe in an offline setting.

Both models use the same architecture and are evaluated after 525 epochs, the highest saved epoch common to both runs. The evaluation uses the same eight source--target windows, raw network weights and NFE20/M8 sampling. The training executions differed in their distributed configuration and gradient accumulation, so the comparison does not isolate the effect of the envelope alone.

The p2 model produces more accurate reconstructions in this comparison. As shown in Table~\ref{tab:prior-envelope}, mean absolute error (MAE) decreases from 0.07699 for p1 to 0.03451 for p2, and CRPS decreases from 0.05836 to 0.02575. The perturbation-to-target Laplacian-variance ratio also decreases substantially, from 65.77 to 1.808, indicating a reduction in the excessive small-scale variability of the generated perturbations.

Coverage, however, decreases from 0.7967 for p1 to 0.6660 for p2. The improved reconstruction and cleaner perturbations therefore do not establish better ensemble calibration.

\begin{table}[!htbp]\centering\small
\caption{Observation-free p1/p2 SR validation after 525 epochs. Both models use raw weights, the same eight validation windows and NFE20/M8 sampling. Scores cover all nine frames and 16 layers on the neural-grid support with 64 meridional rows. MAE, RMSE and CRPS are in physical PV units. Spread/skill is the mean of the window-level ratios. The Laplacian-variance ratio compares generated perturbations with target curvature.}
\label{tab:prior-envelope}
\begin{adjustbox}{max width=\linewidth}
\begin{tabular}{lrrrrrr}
\toprule
Envelope & MAE & RMSE & CRPS & Spread/skill & Coverage90 & Laplacian-variance ratio \\
\midrule
p1 & 0.07699 & 0.10783 & 0.05836 & 1.0117 & 0.7967 & 65.77 \\
p2 & 0.03451 & 0.04720 & 0.02575 & 1.0448 & 0.6660 & 1.808 \\
\bottomrule
\end{tabular}
\end{adjustbox}
\end{table}

\begin{figure}
  \centering
  \includegraphics[width=1.0\textwidth]{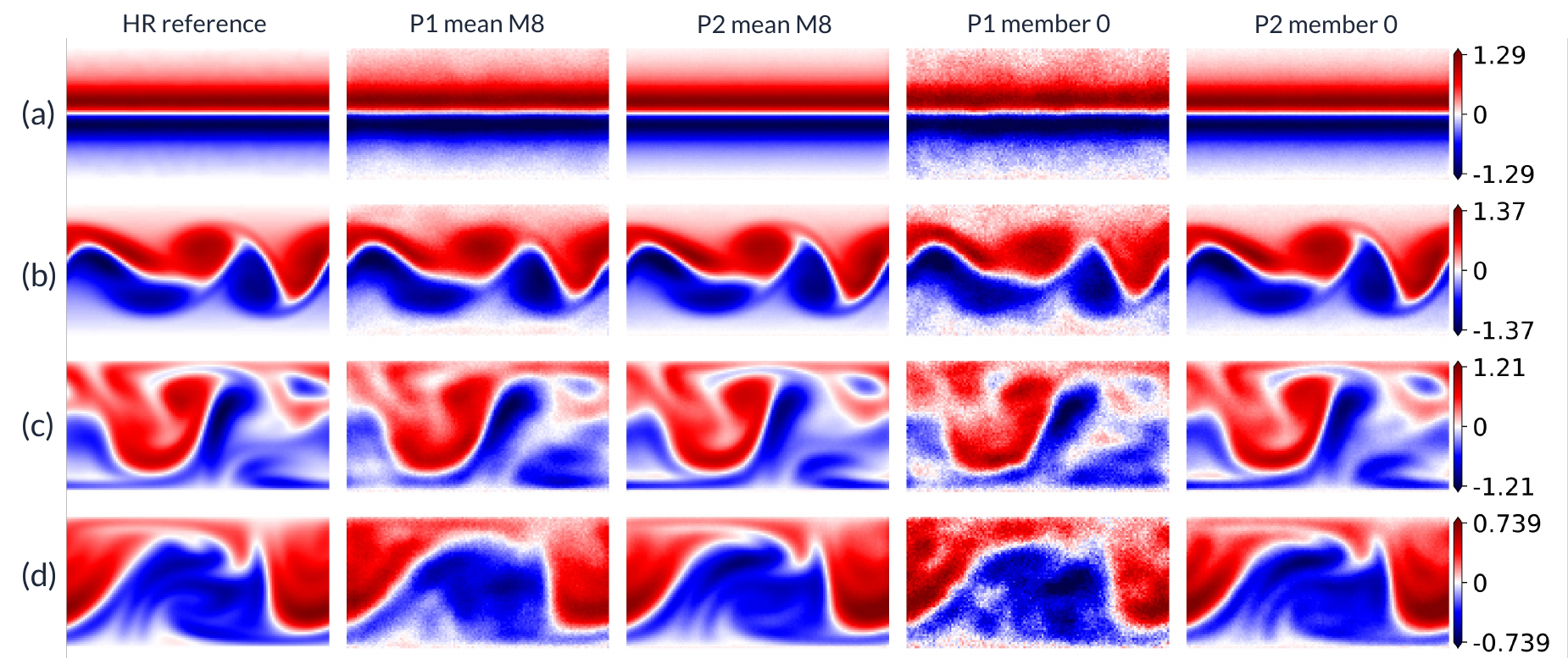} \caption{Observation-free SR reconstructions for four matched validation windows (a)--(d) after 525 epochs. Columns show the HR reference, p1 and p2 ensemble means, and one member from each ensemble. Fields are shown for the center frame and middle layer. Both models use raw weights and NFE20/M8 sampling. PV color ranges are shared across the fields within each row.}
\label{fig:p1-p2-epoch0525-middle}
\end{figure}

Figure~\ref{fig:p1-p2-epoch0525-middle} shows reconstructions for four of the validation windows. For each case, the p2 ensemble mean and individual member exhibit less visible small-scale noise than their p1 counterparts.

\begin{figure}
  \centering
  \includegraphics[width=1.0\textwidth]{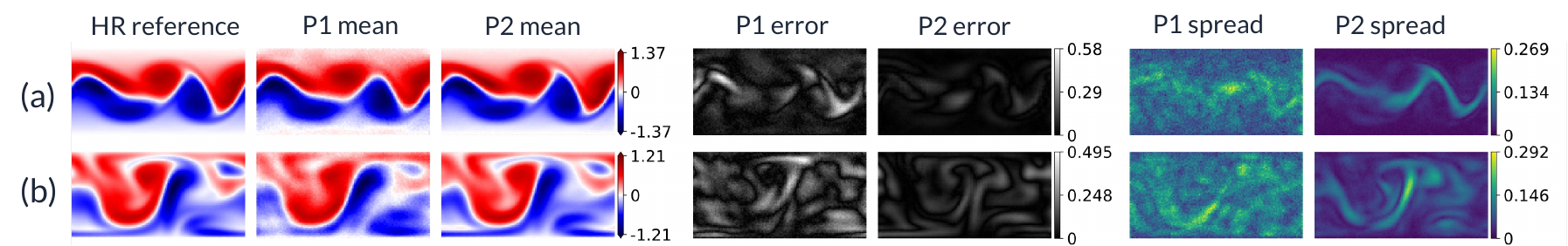} \caption{Observation-free p1 and p2 SR results for two matched validation windows, shown in rows (a) and (b) for the center frame and middle layer. Columns show the HR target, the p1 and p2 ensemble means, their absolute errors and their ensemble standard deviations. Both models use raw weights after 525 epochs and NFE20/M8 sampling. PV, error and spread ranges are shared within each row. The windows come from one validation trajectory.}
\label{fig:prior-envelope-fields}
\end{figure}

Figure~\ref{fig:prior-envelope-fields} shows the reconstruction errors and ensemble spread for two of these windows. The p2 ensemble means have smaller errors, and their spread fields show less granular variability. The common scales make the differences in magnitude visible.

These results support the quadratic-envelope recipe used in this study. They also illustrate that improved reconstruction and smoother spread fields need to be evaluated alongside coverage. The offline SR validation results should be interpreted separately from the recursive narrow-jet assimilation results in Section~\ref{sec:results}.

\subsection{Sampling choices and the scope of the comparison}

Both p1 and p2 validation models use stochastic Heun sampling. The comparison therefore informs our choice of noise-envelope recipe, while the separate effect of the Heun corrector was not measured. The reduced-budget experiments in Section~\ref{sec:results} examine cost and accuracy within the Heun configuration.

The training histories also do not isolate the contribution of continuous stratified time sampling. The results support the p2 recipe in this offline SR setting, while leaving the individual contributions of the sampling and training choices unresolved.

\section{Cycling with nowcast feedback}\label{app:stride}
\FloatBarrier

As shown in Section~\ref{sec:nowcast}, the future frames in each BridgeSRDA window retain useful reconstruction skill over the tested one-unit horizon. This suggests that a nowcast could also provide the initial condition for a subsequent LR forecast. If the resulting analyses remain accurate over repeated cycles, the system could issue observation-conditioned windows less frequently and reduce the cost of neural sampling.

In the main experiments, BridgeSRDA produces a fresh analysis once per physical time unit, and each analysis initializes the next LR forecast. Here we instead issue a fresh analysis every two physical time units and use the one-unit-ahead nowcast for feedback. For example, a window issued at $t=1$ supplies a nowcast at $t=2$. That nowcast is downsampled to initialize the LR forecast used to construct the next window, whose current analysis is issued at $t=3$. We refer to this procedure as stride-two cycling. Observations arriving at the intervening time are not used to revise the nowcast already issued for that time. They can be included as past observations in the next window.

We compare the two procedures on the same two test trajectories using BridgeSRDA-D with NFE20/M10. For each trajectory, normal cycling produces 199 fresh analyses. Stride-two cycling produces 100 fresh analyses and 99 intervening nowcasts, so the number of sampled windows is almost halved. Both procedures are evaluated at the same 199 physical times.

\begin{figure}
  \centering
  \includegraphics[width=1.0\textwidth]{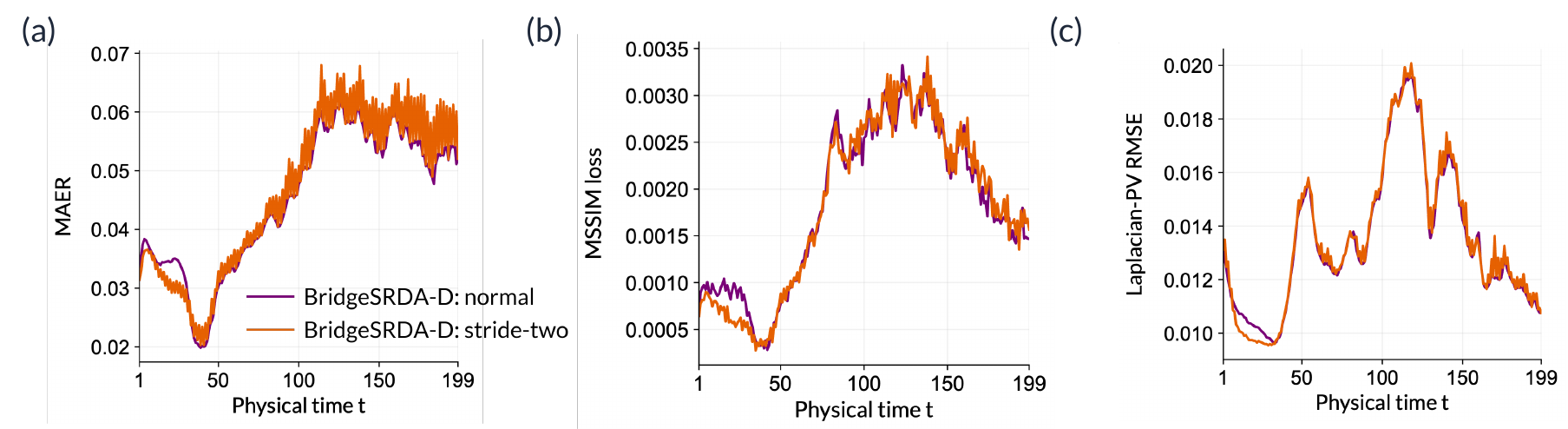} \caption{Reconstruction scores for normal and stride-two BridgeSRDA-D cycling, averaged over two test trajectories: (a) MAER, (b) MSSIM loss and (c) Laplacian-PV RMSE. Both configurations use NFE20/M10. Normal cycling supplies a fresh analysis at every verification time, while stride-two cycling supplies fresh analyses and intervening one-unit-ahead nowcasts. Both are evaluated at the same physical times.}
\label{fig:09-stride2-point}
\end{figure}

\begin{figure}
  \centering
  \includegraphics[width=1.0\textwidth]{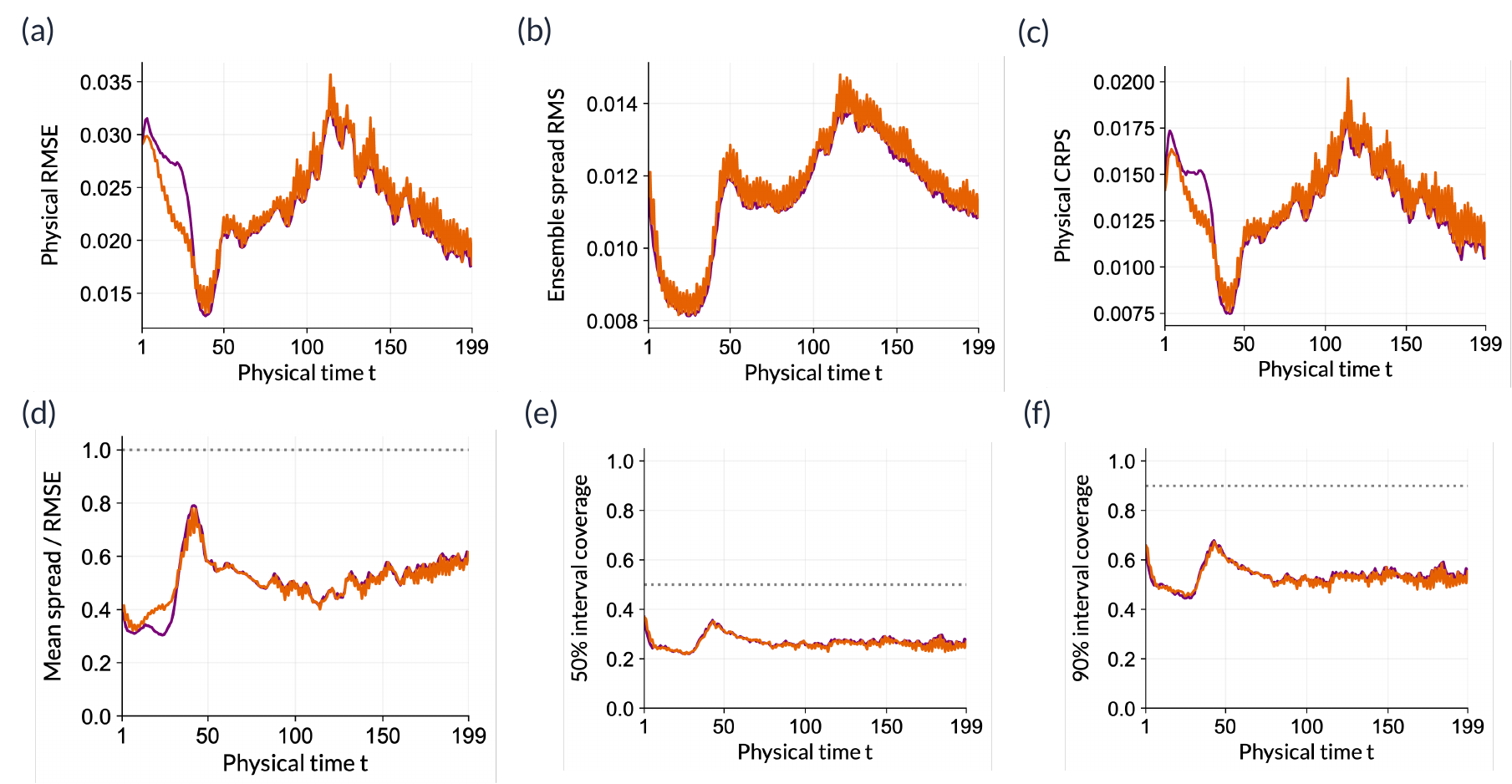} \caption{Ensemble and uncertainty scores for normal and stride-two BridgeSRDA-D cycling, averaged over the same two trajectories as in Figure~\ref{fig:09-stride2-point}. Panels show (a) physical RMSE, (b) RMS ensemble spread, (c) physical CRPS, (d) mean spread/skill, and coverage of the nominal (e) 50\% and (f) 90\% intervals. The line colors are the same as in Figure~\ref{fig:09-stride2-point}. Both configurations use NFE20/M10. The dotted lines indicate a spread/skill ratio of one and nominal interval coverage.}
\label{fig:09-stride2-uq}
\end{figure}

Figures~\ref{fig:09-stride2-point} and \ref{fig:09-stride2-uq} show the scores averaged over the two trajectories. Stride-two cycling reduces the recorded cycling time by 48.1\%, while mean MAER and CRPS increase by 4.1\% and 3.1\%, respectively. Coverage of the nominal 90\% interval decreases by approximately 0.95 percentage points, and both configurations remain underdispersed. During the later period, $t=150$--199, stride-two MAER is approximately 7.1\% higher. The experiment therefore shows a substantial cost reduction with a modest loss of skill on the two trajectories tested.


\begin{figure}
  \centering
  \includegraphics[width=1.0\textwidth]{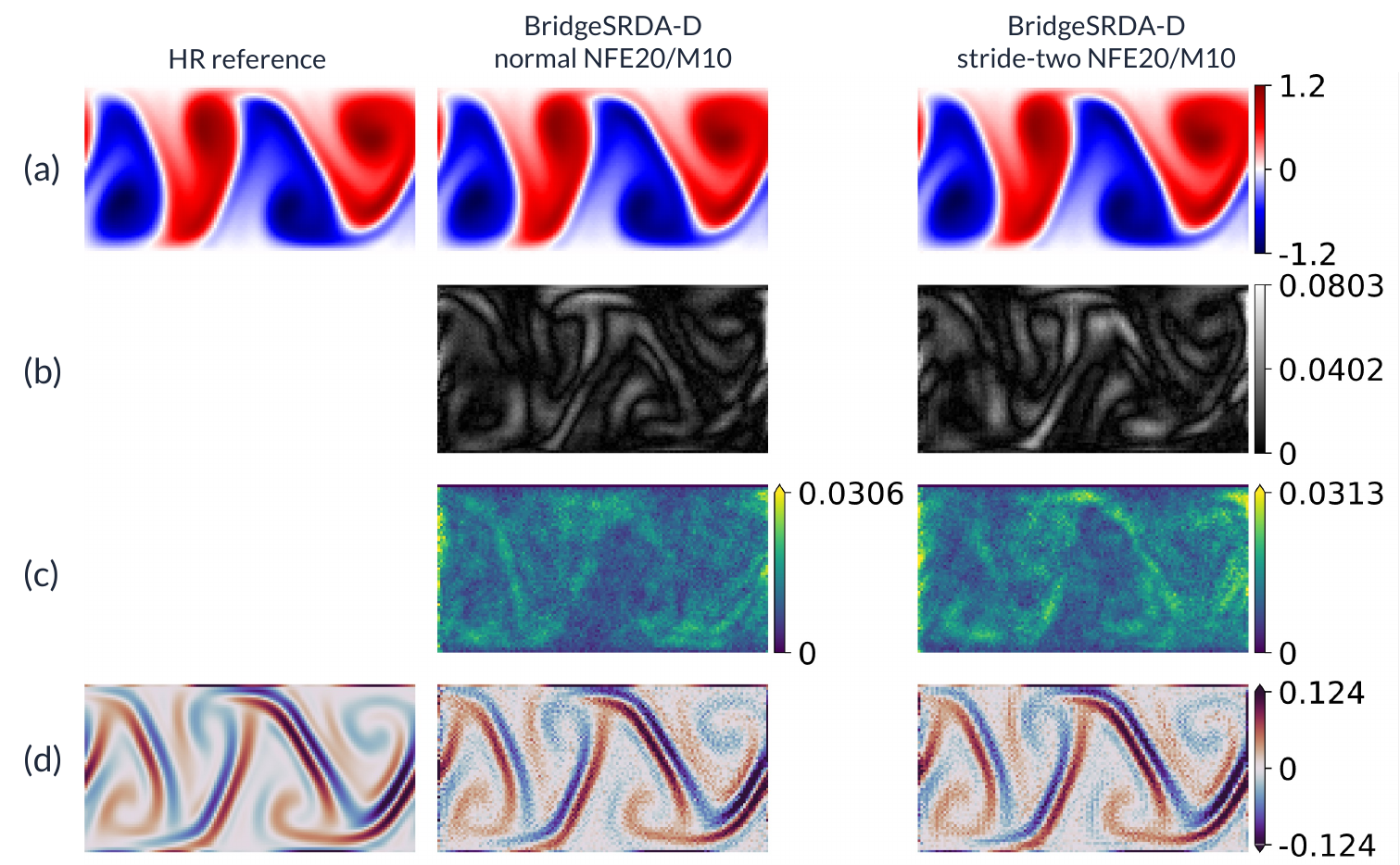} \caption{HR reference and BridgeSRDA-D fields at physical time $t=50$, shown for the middle layer of one test trajectory. Normal cycling supplies a fresh analysis at $t=50$, while stride-two cycling supplies the nowcast issued at $t=49$. Rows show (a) PV, (b) absolute error, (c) ensemble spread and (d) horizontal Laplacian PV. Both BridgeSRDA-D configurations use NFE20/M10. Spread colorbars use method-specific ranges.}
\label{fig:stride-index0200-middle}
\end{figure}

Figure~\ref{fig:stride-index0200-middle} compares the delivered fields at $t=50$ for one test trajectory. Normal cycling supplies a fresh analysis at that time. Stride-two cycling supplies the one-unit-ahead nowcast issued at $t=49$. The two fields retain similar large-scale structure, illustrating the reconstruction obtained at an intervening time when a fresh analysis is not issued.

\FloatBarrier
\section{Additional conditional diffusion experiments}\label{app:diffusion}

Our previous DiffSRDA study used conditional diffusion for probabilistic SRDA with a two-dimensional barotropic jet~\citep{Ayapilla2026DiffSRDA}. Motivated by that result, we first explored conditional diffusion for the three-dimensional dataset considered here.

The diffusion models received noisy HR windows together with LR forecasts and observations. They were trained to predict the added noise using an $L^1$ objective, and sampling reconstructed the full HR state. The training noise schedule contained 1,000 levels. During inference, we used respaced updates from the DDIM family to generate samples with fewer network evaluations~\citep{Song2021DDIM}.

\begin{figure}
  \centering
  \includegraphics[width=1.0\textwidth]{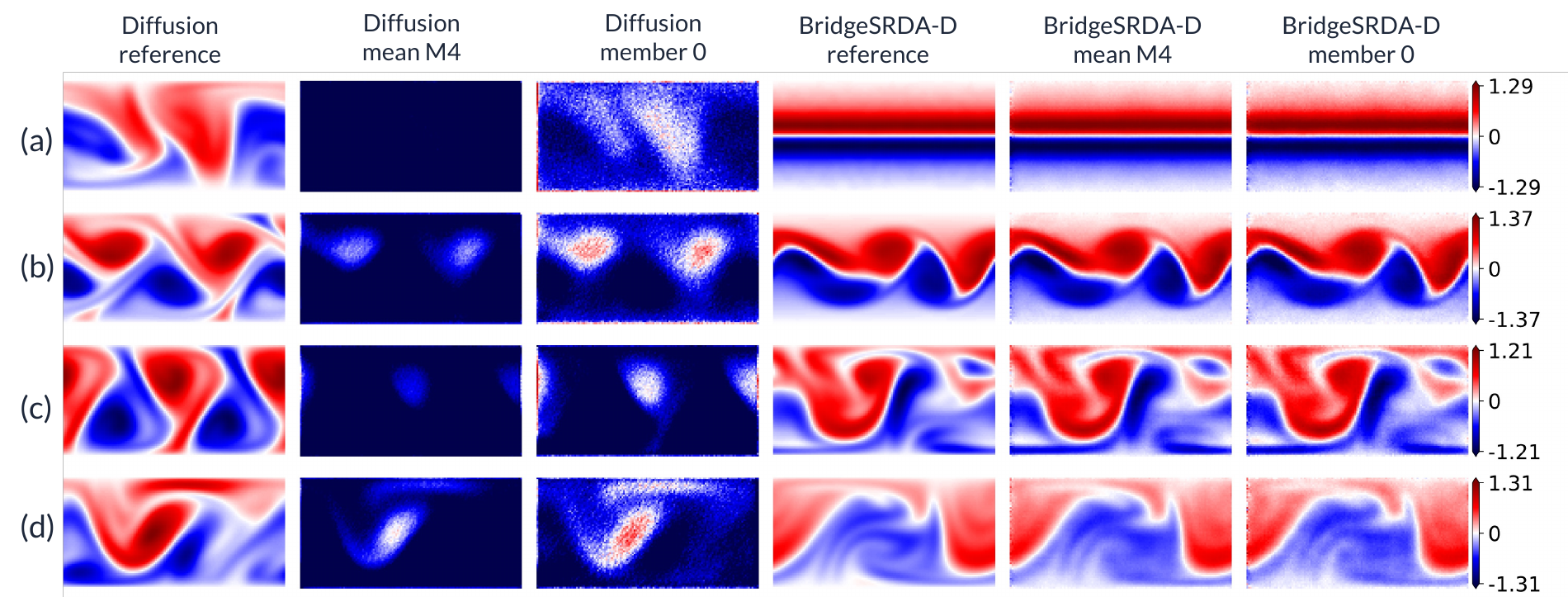} \caption{Validation reconstructions from a conditional diffusion U-Net and BridgeSRDA-D after 125 training epochs. Rows (a)--(d) show four validation examples for each model. For each model, the columns show its HR target, the mean of four generated members and one individual member. Fields are shown for the center frame and middle layer, using non-EMA weights and NFE20. Each model is shown with its own validation target. Rows do not represent matched cases between the two models.}
\label{fig:unet-vs-bridge-ep0125-middle}
\end{figure}

We tested both U-Net and MaxViT architectures. The diffusion U-Net had approximately 89.5 million parameters, a similar count to BridgeSRDA-D. Figure~\ref{fig:unet-vs-bridge-ep0125-middle} compares validation reconstructions from the two models after 125 training epochs. At this stage, BridgeSRDA-D reconstructs coherent flow structures, while the diffusion U-Net's fields remain strongly distorted.

The diffusion U-Net's earlier reconstruction-selected checkpoint recovered some coarse features, but its reconstructions remained inaccurate. The comparison in Figure~\ref{fig:unet-vs-bridge-ep0125-middle} illustrates the behavior of the tested training recipes. Differences in conditioning, training procedures and validation inputs prevent us from attributing the result to the generative formulation alone.

We also examined whether increasing the diffusion sampling budget improved reconstruction. In a separate eight-member MaxViT experiment, increasing the number of evaluations per member from 100 to 500 increased the recorded runtime from approximately 62 to 309 s per window. Over the same change, physical MAE decreased only from approximately 1.30 to 1.25.

The tested diffusion configurations therefore did not provide useful HR reconstruction within the explored training and sampling budgets. These experiments motivated the forecast-informed stochastic transport formulation used in the present study.

\section{Numerical and implementation details}\label{app:implementation}

This appendix collects additional numerical settings and verification details for the experiments described in Section~\ref{sec:setup}.

\subsection{Dynamics and data preparation}

The QG benchmark uses a beta parameter of 0.256, Coriolis parameter of 2.0, maximum jet velocity of 1.0 and vertical sigmoid width of 0.0015625. The spectral hyperdiffusion coefficient is $9.641044626467367\times10^{-6}$, applied to $(k_x^2+k_y^2)^2$, where $k_x$ and $k_y$ are the horizontal spectral wavenumbers. The reduced-gravity parameters are 125.0, 31.25 and 15.625 for the LR, HR and UHR models, respectively, and their numerical integration time steps are 0.01, 0.0025 and 0.00125. LR forecast anchors are saved one physical time unit apart, while HR and UHR frames are saved at intervals of 0.25. All quantities follow the benchmark's nondimensional convention.

The dataset split reconstructed from the available data and saved configurations contains six initial-condition families for training, three for validation and two for testing. Windows from the same trajectory overlap and are not independent realizations. The narrow-jet cycling evaluation uses additional trajectories. To convert the UHR reference to the HR grid, we remove the first meridional row, average-pool by a factor of two in $z/y/x$, and restore a zero first row on the HR grid. Baseline predictions available at UHR are converted in the same way. Neural preprocessing handles the opposite meridional edge: it removes the last row, and output conversion appends a zero last row. All rows are retained when calculating scores. This difference in edge treatment can affect boundary errors and verification ranks, although its sensitivity has not been evaluated separately. Outputs already on the HR grid are not pooled again.

\subsection{Network and sampling implementation}

The observation representations described in Section~\ref{sec:observations} produce 20 input channels for BridgeSRDA-D and 24 for BridgeSRDA-I. Both networks predict nine drift channels, corresponding to the nine physical-time frames. Spatial convolutions use zero padding, so the network itself does not impose periodicity, even though BridgeSRDA-D uses periodic interpolation of observations in the zonal direction. If a dense-observation slot is unavailable, the corresponding source frame is supplied in its place. BridgeSRDA-D and BridgeSRDA-I are trained separately. When sampling budgets are compared, their random increments are coupled using a common Brownian refinement grid.

\subsection{Filter settings and verification conventions}

The HR-LETKF observation-selection cylinder has a full vertical height of 0.0016. The horizontal selection and localization radii given in Section~\ref{sec:comparison-methods} are expressed in domain coordinates. The filter uses an observation-covariance diagonal of 0.022, which is approximately 45.5 times the variance used to generate the observation errors. The effect of this choice has not been evaluated separately.

Rank histograms use strict comparisons, with no randomization of ties. CRPS is calculated using the empirical finite-ensemble score, coverage uses empirical quantiles, and spread/skill is the average of the per-cycle ratios. The main reconstruction table uses the structural-similarity implementation from the original evaluation. For the free-run and sparse-observation comparisons, the structural scores are recomputed using a common CPU implementation for all displayed methods, producing small numerical differences from the main table. For nowcast horizons from 0 to 0.75, the scores use 199 issued windows per trajectory. The horizon-one scores use 198 windows because the final reference state lies beyond the stored sequence.

\section{Execution efficiency and ensemble-budget sensitivity}\label{app:execution-efficiency}

\subsection{Purpose and cycling experiment}

The timings in Section~\ref{sec:results} describe the implementations used for the main ten-trajectory comparison. Here we examine how much deployment cost can be reduced by improving the execution of both BridgeSRDA-D and HR-LETKF on an NVIDIA RTX A6000 GPU. We evaluate complete recursive cycles so that the comparison includes numerical forecasting as well as neural sampling or the filter update.

The experiment uses one narrow-jet trajectory for 199 analyses, with observations incorporated once per physical time unit. The four new runs use the same observation sequence. They consist of optimized HR-LETKF M30 and compiled mixed-precision BridgeSRDA-D at NFE20 with 30, ten and five members. Each run occupies one A6000 GPU, and the runs share the host CPU and filesystem. BridgeSRDA retains the same learned weights, transport construction and member-wise analysis feedback. Each method uses its established initialization procedure. The original BridgeSRDA-D M30 execution for this trajectory provides a reference for assessing the effect of the execution changes.

\subsection{Execution changes and numerical fidelity}

For BridgeSRDA-D, ensemble members are batched and the neural computation is compiled. Neural operations use bfloat16 (BF16) mixed precision, while model parameters, transport states, random increments and Heun updates remain in FP32. NFE20 still consists of ten Heun intervals and 20 network evaluations per member. For HR-LETKF, fixed observation and localization geometry is cached, observations are indexed directly, and the local analysis calculations are batched. The numerical forecast and filter update remain in FP64, and inflation, localization and the observation-covariance convention are unchanged.

All four runs complete the 199 updates without nonfinite outputs or filter fallback blocks. The optimized HR-LETKF ensemble mean closely matches its earlier native-resolution execution, with a maximum absolute difference of approximately $2.4\times10^{-7}$. For BridgeSRDA-D M30, the compiled and original ensemble-mean fields differ by an RMS of approximately 0.0026 in physical PV units, and this difference does not grow systematically over the evaluated cycles. Relative to the original execution, mean MAER increases by approximately 2.7\% and physical RMSE by approximately 1.6\%. These comparisons describe the complete execution configurations, including their numerical differences.

\subsection{Forecasting and update costs}

Table~\ref{tab:execution-costs} separates the cost of numerical forecasting from the cost of the assimilation update. For BridgeSRDA-D, the update column includes the complete neural sampling procedure for all ensemble members. For HR-LETKF, it contains the filter update. The core cycle also includes preparation of the model conditions and feedback of the analysis to the next forecast. It excludes offline training, one-time setup and warm-up, diagnostic scoring and output writing.

\begin{table}[!htbp]\centering\small
\caption{Mean component costs in seconds over 199 cycles on one narrow-jet trajectory. BridgeSRDA-D uses compiled BF16 neural operations and NFE20. HR-LETKF uses the optimized FP64 implementation. Core cycle time includes conditioning and analysis feedback as well as the displayed forecast and update costs. Speed factors are calculated relative to the optimized HR-LETKF M30 core cycle. Offline training, startup, scoring and output writing are excluded.}
\label{tab:execution-costs}
\begin{adjustbox}{max width=\linewidth}
\begin{tabular}{lrrrr}
\toprule
Method & Numerical forecast & Update or sampling & Core cycle & Speed factor \\
\midrule
HR-LETKF M30 & 11.088 & 7.006 & 18.096 & 1.00 \\
BridgeSRDA-D M30 & 0.250 & 5.242 & 5.502 & 3.29 \\
BridgeSRDA-D M10 & 0.240 & 1.859 & 2.104 & 8.60 \\
BridgeSRDA-D M5 & 0.250 & 1.005 & 1.257 & 14.39 \\
\bottomrule
\end{tabular}
\end{adjustbox}
\end{table}

At 30 members, complete BridgeSRDA-D sampling takes approximately 5.24 s, compared with 7.01 s for the optimized LETKF update. The larger difference is in numerical forecasting: the LR forecast takes approximately 0.25 s, while the HR forecast takes 11.09 s. These timings follow the forecast windows required by each method. BridgeSRDA advances its LR source through two physical time units to supply the full reconstruction window, while HR-LETKF advances its HR ensemble through one physical time unit to the next analysis. The latter interval contains four saved output increments, with multiple numerical integration steps within the interval. Combining forecasting and updating, the BridgeSRDA-D M30 core cycle takes 5.50 s, making it approximately 3.3 times faster than the 18.10 s HR-LETKF M30 cycle. Reducing BridgeSRDA-D to ten and five members gives further reductions without changing the observation-refresh rate.

Including diagnostic scoring and output writing gives approximately 18.11 s per analysis for HR-LETKF and 5.51, 2.12 and 1.28 s for BridgeSRDA-D with 30, ten and five members, respectively. These values remain close to the core-cycle times. Compilation and warm-up are one-time startup costs.

\subsection{Reconstruction and ensemble behavior}

\begin{figure}
  \centering
  \includegraphics[width=1.0\textwidth]{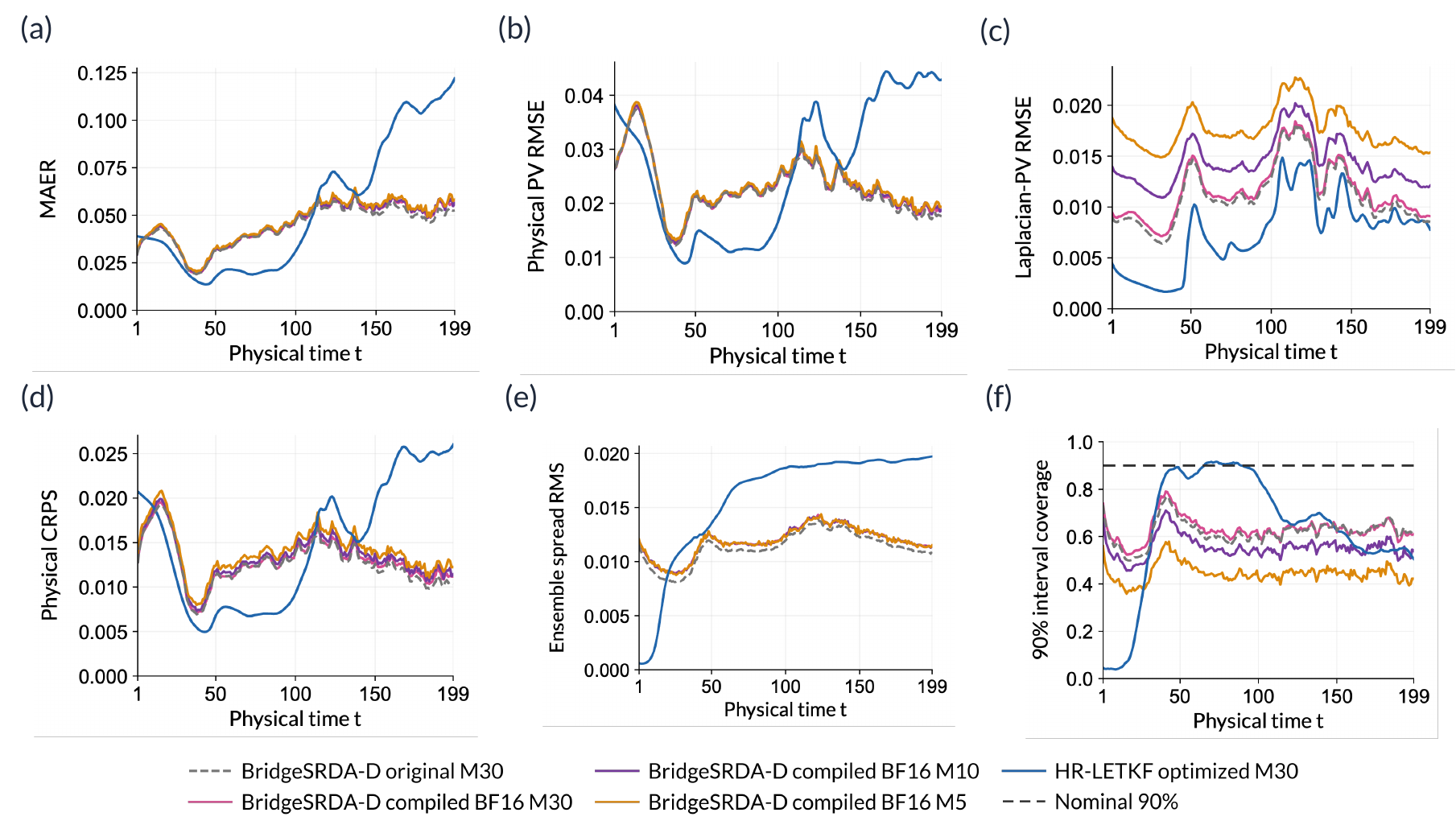} \caption{Reconstruction and uncertainty scores over 199 cycles in the execution-efficiency experiment: (a) MAER, (b) physical PV RMSE, (c) Laplacian-PV RMSE, (d) physical CRPS, (e) RMS ensemble spread and (f) empirical coverage of nominal 90\% intervals. The dashed gray curve is the original BridgeSRDA-D M30 execution. The solid curves show compiled BF16 BridgeSRDA-D with 30, ten and five members and optimized HR-LETKF M30. All BridgeSRDA configurations use the same weights and NFE20. Results are for one test trajectory and are shown without temporal smoothing.}
\label{fig:execution-cycling}
\end{figure}

The lower-cost BridgeSRDA-D configurations retain similar mean reconstruction over the evaluated cycles. Relative to compiled M30, using ten members increases MAER by approximately 1.1\% and physical RMSE by 0.8\%. With five members, the increases are approximately 3.0\% and 2.4\%, respectively. Table~\ref{tab:execution-skill} and Figure~\ref{fig:execution-cycling} show the corresponding scores and their evolution.

\begin{table}[!htbp]\centering\small
\caption{Reconstruction and uncertainty scores averaged over 199 analyses for the same test trajectory. All BridgeSRDA-D configurations use NFE20. ``Original'' denotes the original execution, and the other BridgeSRDA rows use compiled BF16 neural operations. RMSE and CRPS are in physical PV units. Coverage is the empirical coverage of nominal 90\% intervals.}
\label{tab:execution-skill}
\begin{adjustbox}{max width=\linewidth}
\begin{tabular}{lrrrrr}
\toprule
Method & MAER & RMSE & Laplacian RMSE & CRPS & Coverage90 \\
\midrule
BridgeSRDA-D original M30 & 0.04381 & 0.02276 & 0.01121 & 0.01271 & 0.620 \\
BridgeSRDA-D compiled M30 & 0.04501 & 0.02313 & 0.01168 & 0.01295 & 0.630 \\
BridgeSRDA-D compiled M10 & 0.04551 & 0.02333 & 0.01440 & 0.01337 & 0.554 \\
BridgeSRDA-D compiled M5 & 0.04637 & 0.02368 & 0.01766 & 0.01405 & 0.445 \\
HR-LETKF optimized M30 & 0.05327 & 0.02663 & 0.00744 & 0.01489 & 0.638 \\
\bottomrule
\end{tabular}
\end{adjustbox}
\end{table}

The effect of ensemble size is more noticeable in the fine-scale and uncertainty diagnostics. Laplacian RMSE increases as the ensemble is reduced, and empirical coverage decreases from 0.630 for compiled M30 to 0.554 for M10 and 0.445 for M5. The domain-wide RMS spread remains similar, but the spread maps become more granular with fewer members. A smaller ensemble also provides fewer samples from which to estimate percentile intervals, particularly in the tails. Finite-ensemble effects can therefore affect the measured coverage. The experiments do not isolate those effects from changes in the recursively evolving forecasts. The results show that a similar mean PV reconstruction can be retained at lower cost, while fine-scale accuracy and the resolution of the sampled uncertainty require additional attention.

\begin{figure}
  \centering
  \includegraphics[width=1.0\textwidth]{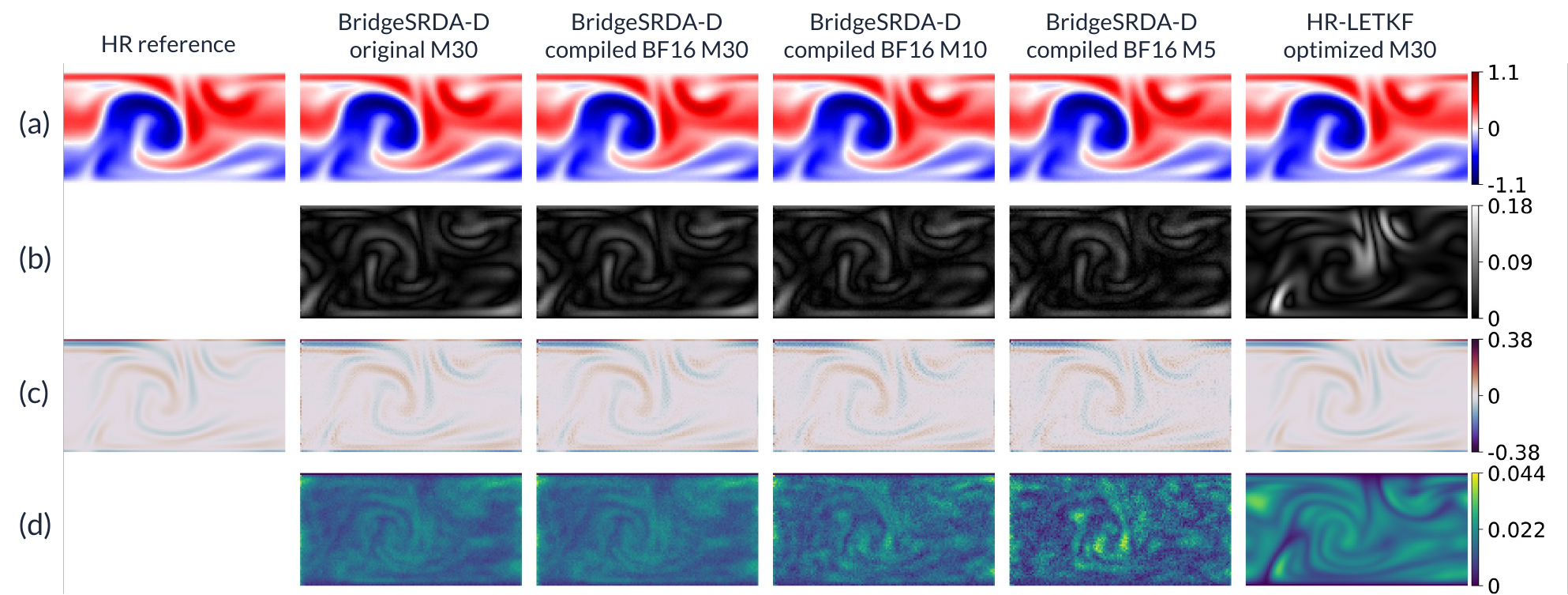} \caption{Middle-layer fields at physical time $t=125$ for the test trajectory used in the execution-efficiency experiment. Columns show the HR reference, original BridgeSRDA-D M30, compiled BF16 BridgeSRDA-D with 30, ten and five members, and optimized HR-LETKF M30. Rows show (a) ensemble-mean PV, (b) absolute error, (c) horizontal Laplacian PV and (d) ensemble standard deviation. All BridgeSRDA configurations use NFE20. Color limits are shared across methods within each row: PV $\pm1.1$, error 0--0.18, Laplacian PV $\pm0.38$ and spread 0--0.044. The Laplacian uses the verification stencil defined in Section~\ref{sec:metrics}. Reference error and spread cells are blank.}
\label{fig:execution-fields}
\end{figure}

Figure~\ref{fig:execution-fields} illustrates the reconstructed fields at $t=125$ in the middle layer. The original and compiled M30 PV fields are very similar, and the M10 and M5 means retain the principal flow structures. Differences are more apparent in the Laplacian and spread fields. At this time, HR-LETKF has smoother spread and more accurate curvature, while its absolute-error field shows larger spatially coherent reconstruction errors.

For this trajectory and hardware, the experiment demonstrates two sources of computational flexibility: reducing the cost of execution and changing the number of sampled members. The appropriate choice depends on whether the application prioritizes mean reconstruction, fine-scale detail or probabilistic information.

\end{document}